\documentclass[journal]{IEEEtran}
\usepackage{graphics}
\usepackage{mathrsfs}
\usepackage{color}
\usepackage{amsfonts}
\usepackage{color}
\usepackage{float}
\usepackage{cite}
\usepackage{graphicx}
\usepackage{amsmath}
\usepackage{amsthm}
\usepackage{multicol}
\usepackage{multirow}
\usepackage{color}
\usepackage{ifpdf}
\usepackage{flushend}
\usepackage{txfonts}
\usepackage{mdwlist}

\theoremstyle{definition} \newtheorem{remark}{Remark}
\theoremstyle{definition} 
\theoremstyle{definition} \newtheorem{theorem}{Theorem}
\theoremstyle{definition} \newtheorem{lemma}{Lemma}
\theoremstyle{assumption} 
\theoremstyle{definition} \newtheorem{definition}{Definition}
\theoremstyle{definition} 
\theoremstyle{definition} \newtheorem{example}{Example}
\theoremstyle{definition} \newtheorem{corollary}{Corollary}
\theoremstyle{definition} 
\theoremstyle{definition}

\usepackage{ifpdf}
\ifpdf
  \usepackage{epstopdf}
\fi

\newcommand{\R}{\mathbb{R}}
\newcommand{\Z}{\mathbb{Z}}

\newcommand{\setS}{\mathcal{S}}

\newcommand{\Int}{\text{Int}}
\newcommand{\e}{\text{e}}

\newcommand{\X}{\mathcal{X}}

\newcommand{\T}{\text{T}}
\newcommand{\cl}{\text{cl}}

\begin{document}

\title{
Safety Margins of Inverse Optimal ISSf Controllers
}

\author{Ziliang~Lyu,~
        Yiguang~Hong,~
        Lihua~Xie,~
        and~Miroslav Krstic

\thanks{Z. Lyu and L. Xie are with the School of Electrical, and Electronic Engineering, Nanyang Technological University, Singapore 639798, Singapore (e-mail: ziliang.lyu@ntu.edu.sg; elhxie@ntu.edu.sg)}

\thanks{Y. Hong is with the Department of Control Science and Engineering, Tongji University, Shanghai 201804, China (e-mail: yghong@iss.ac.cn)}

\thanks{M. Krstic is with the Department of Mechanical and Aerospace Engineering, University of California at San Diego, La Jolla, CA, 92093-0411 USA (e-mail: mkrstic@ucsd.edu)}
}

\maketitle

\begin{abstract}
We investigate the gain margin of a general nonlinear system under an inverse optimal input-to-state safe (ISSf) controller of the form $u=u_0(x)+\bar{u}^*(x,u_0)$, where $u_0$ is the nominal control and $\bar{u}^*$ is the inverse optimal safety filter that minimally modifies the nominal controller's unsafe actions over the infinite horizon. By first establishing a converse ISSf-BF theorem, we reveal the equivalence among the achievability of input-to-state safety by feedback, the achievability of inverse optimality, and the solvability of a Hamilton-Jacobi-Isaacs equation for an ISSf control barrier function (ISSf-CBF), which is the value function of inverse optimal ISSf gain assignment. We develop a collection of safety margin results on the overall inverse optimal safe control $u=u_0+\bar{u}^*$. In the absence of disturbances, we find that standard inverse optimal safe controllers have a certain degree of gain margin depending on whether the open-loop vector field $f(x)$ or the nominal controller $u_0$ acts safely along the entire safety boundary. Specifically, when $f(x)$ acts safely but $u_0$ acts unsafely, the gain on the overall control $u=u_0+\bar{u}^*$ can be decreased by up to half; and when $f(x)$ acts unsafely, we establish that, if $u_0$ acts safely, the gain can be increased arbitrarily, whereas if $u_0$ acts unsafely, the overall control recovers the full gain margin $[1/2,\infty)$. It is shown, however, that under control gain variation, the safe set of these controllers is locally asymptotically stable, which implies that their safety is sensitive to large but bounded disturbances. To make inverse optimal ISSf controllers robust to gain variation, we propose a gain margin improvement approach at the expense of an increased control effort. This improvement allows the inverse optimal safe control to inherit the standard gain margin of $[1/2,\infty)$ without requiring prior knowledge of whether $f(x)$ or $u_0$ acts safely on the safety boundary, while simultaneously ensuring global asymptotic stability of the resulting safe set. In the presence of disturbances, this improvement idea renders inverse optimal ISSf controllers robust to gain variations with the same gain margin of $[1/2,\infty)$.
\end{abstract}


\section{Introduction}

\subsection{Safety and Barrier Functions}

Safety is a fundamental requirement in control systems \cite{tomlin1998conflict, ames2019control, wabersich2023data}. The past two decades have witnessed the rise of barrier functions as one of the most promising tools for safety verification and control \cite{prajna2007framework, ames2016control}. Barrier functions are essentially Lyapunov-like methods and thus inherit the advantage of avoiding the explicit computation of solutions or finite abstractions of nonlinear systems. Recent results suggest that many Lyapunov-theoretic techniques in stability analysis and stabilization design can be suitably adapted to safety considerations; see, e.g., \cite{wieland2007constructive, ames2014control, xu2015robustness, romdlony2016new, kolathaya2018input, glotfelter2020nonsmooth, lyu2022small, krstic2023inverse, abel2023prescribed, wu2025optimization, dong2025robust, wu2025continuous}. In \cite{prajna2007framework}, it is shown that the barrier certificate condition $\dot{h}(x)\geq0$ is sufficient for the forward invariance of a safe set $\setS=\{x\in\R^n:h(x)\geq0\}$ that does not interset the unsafe region. However, this condition is overly conservative, as it requires all super-level set of the barrier function to be forward invariant. A major breakthrough in addressing this limitation is the zeroing barrier function (ZBF) condition, $\dot{h}(x)\geq-\alpha(h(x))$, first proposed in \cite{kong2013exponential} with a linear decay rate $\alpha(s)=\lambda s$ and later generalized in \cite{xu2015robustness} to the case of nonlinear decay rates, where $\alpha$ is a $K$-function. Under ZBF conditions, only one super-level set of the barrier function is required to be forward invariant, and the safe set is asymptotically stable, which implies that any state trajectory initialized outside the safe set will eventually reach the safe set. For the converse ZBF problem, \cite{lyu2023converseZBF , ames2016control} showed that the existence of a ZBF is also necessary for safety. In \cite{nguyen2016exponential, xu2018constrained, xiao2021high, tan2021high, lyu2023highsmall, abel2023prescribed}, ZBFs have been extended to handle high-relative-degree safety constraints. Interestingly, an idea analogous to high-relative-degree ZBFs has appeared in the field of nonvershooting control \cite{krstic2006overshoot} in 2006. Parallel to the extension of Lyapunov functions to control Lyapunov functions (CLFs), control barrier functions (CBFs) were first introduced in \cite{wieland2007constructive} based on barrier certificates and later refined with ZBFs in \cite{xu2015robustness}. In \cite{ames2014control}, quadratic programs (QPs) were introduced to unify CLFs and CBFs, enabling local Lipschitz control synthesis.

In practical problems, it is desirable that for a safe system,
\begin{itemize}
   \item the safety is not sensitive to disturbance inputs;
   \item the ``cost'' of maintaining safety is minimized.
\end{itemize}
To achieve these objectives, input-to-state safety (ISSf) \cite{kolathaya2018input, lyu2022small, krstic2006overshoot} and inverse optimal safe control \cite{krstic2023inverse} play a crucial role.

\subsection{Input-to-State Safety}

ISSf characterizes the dependence of system safety on the magnitude of disturbance inputs. It is an extension of input-to-state stability (ISS) \cite{sontag1995characterizations}. Intuitively, the ISSf property implies that safety violations caused by disturbance inputs remain bounded, provided that these inputs are bounded. Though the ISSf concept was formally introduced to the CBF community by \cite{romdlony2016new, kolathaya2018input}, this notion originally appeared in 2006 in the nonovershooting control context \cite{krstic2006overshoot}. Compared with the notion of robust safety \cite{jankovic2018robust , liu2021converse}, the ISSf does not require the disturbance input to take value in a known compact set. In \cite{kolathaya2018input, lyu2022small, krstic2023inverse}, it was shown that the existence of an ISSf barrier function (ISSf-BF) is a sufficient condition for ISSf. Recently, ISSf has attracted a lot of attentions in the safety-critical control community. In \cite{lyu2022small, lyu2023highsmall}, ISSf-based small-gain theorems were developed for the safety verification of interconnected systems. In \cite{alan2023control}, ISSf was used to capture the impact of disturbances in the robust design of a connected automated vehicle. As AI technologies are increasingly applied in control, ISSf or similar concepts have been used to characterize the influence of learning errors on the safety of closed-loop systems \cite{fisac2018general, Compton2025Learning}.

\subsection{Inverse Optimal Safe Control}

Optimality is one of the most important properties for control systems. The main research effort in the history of optimal control theory focuses on the connection between optimality and stability. An appealing advantage of optimal control is the guaranteed robustness with respect to variations in the plant. For linear systems, it is well-known that optimal controllers resulting from a quadratic criterion possess an $[1/2, \infty)$ gain margin \cite{anderson2007optimal, safonov1977gain}. Similar robustness properties have been shown to hold also for nonlinear control systems \cite{glad1987robustness, tsitsiklis1984guaranteed}, in which the control penalty is relaxed to be a positive definite function. However, the optimal control for nonlinear systems is hampered by the difficulty of solving Hamilton-Jacobi-Isaacs (HJI) partial differential equations. Different form the direct approach, the inverse optimal problem designs a controller and then constructs a cost functional with the information of the controller and system dynamics. Inverse optimal controllers inherit gain margins guaranteed by optimality, while avoiding HJI computations. The first inverse optimal control problem was posed and solved by Kalman \cite{kalman1964linear} in 1960s. Over the six decades, inverse optimal stabilization has been widely studied; see, e.g., \cite{kalman1964linear, anderson2007optimal, molinari1973stable} for linear systems and \cite{moylan1973nonlinear, freeman1996inverse, sepulchre2012constructive, krstic1998inverse, li1997optimal, lu2024small} for nonlinear systems.

Compared to stability, there are few results on the connection between safety and inverse optimality. The first inverse optimal safe control result appeared in the recent paper \cite{krstic2023inverse}, which showed that, if the nonlinear control system $\dot{x}=f(x)+g_1(x)w+g_2(x)u$ has an ISSf-CBF $h(x)$, then there is a matrix-valued function $R(x,u_0)=R(x,u_0)^T>0$ such that the inverse optimal ISSf controller $u=u_0+\bar{u}^*(x,u_0)=u_0+2R(x,u_0)^{-1}(L_{g_2}h)^T$ optimizes the cost functional in the form of
\begin{equation}\label{eq:intro-cost}
    \sup_u\inf_w\int_0^\infty\Big[l(x,u_0)-(u-u_0)^\T R(x,u_0)(u-u_0)-\gamma(|w|)\Big]dt
\end{equation}
where $l(x,u_0)$ penalizes unsafe actions, $u_0$ is the nominal controller, $\bar{u}^*(x,u_0)$ is the override controller used to modify the control input that violates safety, and $\gamma$ is a class $K_\infty$ function. To avoid confusions, we refer to $u=u_0+\bar{u}^*(x,u_0)$ as the overall controller. Intuitively, the cost functional (\ref{eq:intro-cost}) rewards both safety and close adherence to the nominal control, while penalizing the disturbance when it reduces safety with large energy. In fact, many safety-critical controllers are inverse optimal with respect to (\ref{eq:intro-cost}), including CBF-QP controllers \cite{ames2016control} and Sontag-type CBF controllers \cite{krstic2023inverse}.

\subsection{Our Problem and Motivation}

Although \cite{krstic2023inverse} has established a relatively comprehensive inverse optimal safe control framework, two fundamental problems remain unsolved.
\begin{itemize}
   \item First, is there any connection between ISSf and inverse optimality? This problem differs from the one studied in \cite{krstic2023inverse}, which reveals the connection between CBFs and inverse optimality. This problem is motivated by the fact that, under some observability or detectability assumptions, there exists an equivalent relationship among stability, passivity, and inverse optimality, which has been utilized to handle unmodeled input dynamics in stabilization \cite{sepulchre2012constructive, krstic1998inverse}. Returning to safety, it is natural to expect that one can reveal the connection between ISSf and inverse optimality.
   \item Second, does an inverse optimal ISSf controller of the form $u= u_0+\bar{u}^*(x,u_0)$ inherit the $[1/2,\infty)$ gain margin of inverse optimal stabilizers? This question is motivated by the fact that pure ISSf controllers can handle additive disturbances in the control input; however, variations in the control gain fall into the category of multiplicative uncertainties, which a pure ISSf controller cannot address. It is natural to expect that inverse optimal ISSf controllers can inherit the gain margin property of inverse optimal stabilizers, thereby ensuring robustness against uncertainties in the control gain. We remark that, in the special case where $u_0=0$, the overall control $u=u_0+\bar{u}^*(x,u_0)$ reduces to $u= \bar{u}^*(x,0)$, on which the control penalty is quadratic, and thus it is natural for such a controller to inherit the $[1/2,\infty)$ gain margin. However, when $u_0$ is not zero, the penalty on the overall control $u=u_0+\bar{u}^*(x,u_0)$ is neither quadratic nor positive definite. As a result, the standard analysis in optimal control theory \cite{anderson2007optimal} cannot be directly applied to prove the gain margin of our problem.
\end{itemize}

\subsection{Contributions}

Our contributions are summarized as follows.
\begin{itemize}
  \item To reveal the connection between ISSf and inverse optimality, we establish a converse ISSf-BF theorem to show that the existence of an ISSf-BF is not only a sufficient but also a necessary condition for ISSf (see Theorem \ref{thm:converse-ISSF-BF}). In contrast to the sufficiency results in \cite{kolathaya2018input, lyu2022small, krstic2023inverse, lyu2023highsmall}, our result is both necessary and sufficient.
  \item With the converse ISSf-BF theorem, we show that the existence of a safety-critical controller that renders a system ISSf is equivalent to the solvability of an HJI equation associated with a meaningful inverse optimal ISSf gain assignment problem (see Theorem \ref{thm:inverse-optimal}). In contrast to \cite{krstic2023inverse}, which established the connection between CBFs and inverse optimality, our result focuses on the equivalence between ISSf controllers and inverse optimality.
  \item We study whether inverse optimal ISSf controllers of the form $u=u_0+\bar{u}^*(x,u_0)$ inherit the well-known gain margin property of $[1/2,\infty)$ in conventional optimal stabilization \cite{anderson2007optimal, safonov1977gain, glad1987robustness, tsitsiklis1984guaranteed, sepulchre2012constructive, krstic1998inverse}. Our findings reveal that, in the absence of disturbances, inverse optimal safe controllers derived from standard CBF conditions \cite{krstic2023inverse, ames2016control} have a certain degree of gain margin; however, this margin is not necessarily $[1/2,\infty)$ (see Theorem \ref{thm:krstic-gain-margin} and Corollary \ref{cor:CBF-GM}). Specifically, for the case where, along the safety boundary, the open-loop vector field $f(x)$ acts safely but the nominal control $u_0$ acts unsafely, the gain on the overall inverse optimal safe control $u=u_0+\bar{u}^*$ can be decreased by up to half. For the case where, along the safety boundary, $f(x)$ acts unsafely, we establish that, if $u_0$ acts safely, the gain can be increased arbitrarily, whereas if $u_0$ acts unsafely, the overall control recovers the standard gain margin of $[1/2,\infty)$. Nevertheless, the safe set of these controllers is locally asymptotically stable when the control gain is variant, which implies that their safety is sensitive to large but bounded disturbances.
  \item To make inverse optimal ISSf controllers robust to variations in control gain, we further propose a gain margin improvement approach. By increasing the control effort, the inverse optimal safe controllers can inherit the standard gain margin of $[1/2,\infty)$ without requiring prior knowledge of whether the open-loop vector field or the nominal controller acts safely on the safety boundary, while simultaneously ensuring that the resulting closed-loop system is globally asymptotically stable with respect to the safe set (see Theorem \ref{thm:GM-inv-opt-ctrl-dom-zero-dist}). This idea is subsequently used to improve inverse optimal ISSf controllers, enabling them to possess gain margin $[1/2,\infty)$ (see Theorem \ref{thm:GM-domination}).
\end{itemize}

\subsection{Notions and Terminologies}

Throughout this paper, `$\circ$' denotes the composition operator, i.e., $f\circ{g}(s)=f(g(s))$; `T' denotes the transpose operator; $\alpha'(s)$ denotes the derivative of a continuously differentiable function $\alpha$ with respect to $s$; $\R$ and $\Z$ denote the set of real numbers and integers, respectively; $\R_{\geq0}$ and $\Z_{\geq0}$ denote the set of nonnegative real numbers and nonnegative integers, respectively. Given a closed set $\setS$, denote by $\partial\setS$ the boundary of $\setS$. For any $x$ in Euclidean space, $|x|$ is its norm, and $|x|_{\setS}=\inf_{\bar{x}\in\setS}|x-\bar{x}|$ denotes the point-to-set distance from $x$ to the set $\setS$. Denote by $L_\infty^m$ the set of essentially bounded measurable functions $u:\R_{\geq0}\rightarrow\R^m$. For any $u\in L_\infty^m$, $\|u\|_J$ stands for the supremum norm of $u$ on an interval $J\subseteq\R_{\geq0}$ (i.e., $\|u\|_{J}=\sup_{t\in J}|u(t)|$), and we take $\|u\|=\|u\|_{[0,\infty)}$ for simplicity.

A continuous function $\gamma$: $\mathbb{R}_{\geq0}\rightarrow\mathbb{R}_{\geq0}$ with $\gamma(0)=0$ is of class $K$ ($\gamma\in K$), if it is strictly increasing. A function $\gamma\in{K}$ is of class $K_\infty$ ($\gamma\in K_\infty$) if it is unbounded. A function $\beta:\R_{\geq0}\times\R_{\geq0}\rightarrow\R_{\geq0}$ is of class $KL$ ($\beta\in KL$), if for each fixed $t$, the mapping $s\mapsto\beta(s,t)$ is of class $K$, and for each fixed $s\geq0$, $t\mapsto\beta(s,t)$ is decreasing to zero as $t\rightarrow+\infty$. For safety analysis, we introduce the following extended comparison functions accordingly. A continuous function  $\gamma:\R\rightarrow\R$ with $\gamma(0)=0$ is of extended class $K$ ($\gamma\in EK$) if it is strictly increasing. In particular, a function $\gamma\in{EK}$ is of extended class $K_\infty$ ($\gamma\in EK_\infty$) if it is unbounded.

\section{Preliminaries}

\subsection{Safety and Barrier Functions}

Consider the system
\begin{flalign}\label{eq:zero-disturbance-system}
    \dot{x}=f(x)
\end{flalign}
where $x\in\R^n$ is the state and $f:\R^n\rightarrow\R^n$ is locally Lipschitz continuous. Denote by $x(t,x_0)$ the solution of (\ref{eq:zero-disturbance-system}) starting from initial state $x(0)=x_0$.

Denote by $\X_u$ the unsafe region of system (\ref{eq:zero-disturbance-system}). Let $\setS$ be a closed set such that $\setS\bigcap\X_u=\emptyset$. In the barrier function literature \cite{prajna2007framework, ames2016control}, $\setS$ is characterized by a continuously differentiable function $h(x)$:
\begin{subequations}\label{eq:safety-and-bf-amespaper}
\begin{flalign}
    \setS&=\{x\in\R^n:h(x)\geq0\},
     \\
    \partial\setS&=\{x\in\R^n:h(x)=0\},
     \\
     \Int(\setS)&=\{x\in\R^n:h(x)>0\}.
\end{flalign}
\end{subequations}
As indicated in \cite[Lemma 1]{lyu2023converseZBF}, (\ref{eq:safety-and-bf-amespaper}) is equivalent to
\begin{subequations}\label{eq:barrier-radial-unbounded}
\begin{flalign}
    &h(x)=0,\;\;\forall x\in \partial \setS,
     \label{eq:barrier-radial-unbounded-A}\\
    &\alpha_1(|x|_{\cl(\R^n\backslash\setS)})\leq h(x)\leq \alpha_2(|x|_{\cl(\R^n\backslash\setS)}),\;\;\;\;\forall x\in\Int(\setS),
     \label{eq:barrier-radial-unbounded-B}\\
    &\alpha_2(-|x|_\setS)\leq h(x)\leq \alpha_1(-|x|_\setS),\;\;\;\;\forall x\in \R^n\backslash\setS
      \label{eq:barrier-radial-unbounded-C}
\end{flalign}
\end{subequations}
where $\alpha_1$ and $\alpha_2$ are $EK$-functions.

\begin{definition}[Safety \cite{prajna2007framework, ames2016control}]
System (\ref{eq:zero-disturbance-system}) is safe on a set $\setS$ if, for any $x_0\in\setS$, $x(t,x_0)\in\setS$ for all $t\geq0$.
\end{definition}

\begin{definition}[Set Asymptotic Stability \cite{hahn1967stability}]
A closed set $\setS$ is said to be locally asymptotically stable (LAS) for system (\ref{eq:zero-disturbance-system}) if
\begin{flalign}
    |x(t,x_0)|_\setS\leq\beta(|x_0|_\setS,t),\;\;\;\;\forall x_0\in D,\;\;\forall t\geq0
\end{flalign}
where $\beta\in KL$ and $D$ is an open set such that $\setS\subset D\subseteq\R^n$. Particularly, $\setS$ is said to be globally asymptotically stable (GAS) if $D=\R^n$. In stability analysis, $D$ is referred to as the region of attraction.
\end{definition}

\begin{definition}[Zeroing Barrier Function \cite{xu2015robustness, ames2016control}]
A continuously differentiable function $h:D\rightarrow\R$ satisfying (\ref{eq:safety-and-bf-amespaper}) is a zeroing barrier function (ZBF) if
\begin{flalign}\label{eq:ZBF-def}
    L_fh(x)\geq-\alpha(h(x)),\;\;\;\;\forall x\in D
\end{flalign}
where $\alpha\in EK$ and $D$ is an open set with $\setS\subset D\subseteq \R^n$.
\end{definition}

\begin{lemma}[\cite{xu2015robustness, ames2016control, lyu2023converseZBF}]\label{lemma:zbf-safety}
System (\ref{eq:zero-disturbance-system}) is safe on $\setS$ and LAS with respect $\setS$ if and only if it has a ZBF $h(x)$ satisfying (\ref{eq:safety-and-bf-amespaper}) and (\ref{eq:ZBF-def}). In particular, the safe set $\setS$ is GAS if $D=\R^n$.
\end{lemma}

The ZBF condition (\ref{eq:ZBF-def}) only requires only the zero super-level set of $h(x)$ is forward invariant. This condition is less conservative than the barrier certificate condition \cite{prajna2007framework, prajna2007convex, wieland2007constructive}
\begin{flalign}\label{eq:BC-cond}
    L_fh(x)\geq0,\;\;\forall x\in D,
\end{flalign}
which requires all super-level sets of $h(x)$ to be forward invariant. As shown in the following example, a safe system may not have a barrier certificate, even though it is a linear scalar system.

\begin{example}
Consider the linear scalar system $\dot{x}=-x$ with the safe region $\setS=\{x:x\geq0\}$. Clearly, this system is safe. Now we show that it is impossible for this system to have a barrier certificate $h(x)$ satisfying (\ref{eq:safety-and-bf-amespaper}) and (\ref{eq:BC-cond}) by contradiction. Assume that this is not true. Differentiating $h(x)$ along $\dot{x}=-x$ yields $\dot{h}(x)=-\nabla h(x)x$, where $\nabla h(x)$ is the gradient of $h(x)$ with respect to $x$. Because $h(x)$ is continuously differentiable, it follows from the barrier function candidate definition (\ref{eq:safety-and-bf-amespaper}) that there is a sufficiently small $\delta>0$ such that $\nabla h(x)>0$ for all $x\in(0,\delta)$. This implies that, for all $x\in(0,\delta)$, $\dot{h}(x)=-\nabla h(x)\cdot x<0$, contradicting (\ref{eq:BC-cond}). Therefore, it is impossible for the system $\dot{x}=-x$ to have a barrier certificate satisfying (\ref{eq:safety-and-bf-amespaper}) and (\ref{eq:BC-cond}).
\end{example}

\subsection{Input-to-State Safety}

Now we introduce the input-to-state safety (ISSf) that can be used to characterize the effect of disturbance on safety. Consider the system
\begin{flalign}\label{eq:iss-system}
    \dot{x}=f(x,w)
\end{flalign}
where $x\in\R^n$ is the state, $w\in L_\infty^m$ is the external input, and $f:\R^n\times\R^m\rightarrow\R^n$ is a locally Lipschitz continuous. Denote by $x(t,x_0,w)$ the solution of (\ref{eq:iss-system}) with the initial condition $x(0)=x_0$ and external input $w\in L_\infty^m$.

Due to $w$, the state trajectory $x(t,x_0,w)$ may leave the set $\setS$ defined in (\ref{eq:safety-and-bf-amespaper}). Introduce a larger set
\begin{flalign}\label{eq:setS-w-new-def}
    \setS_w=\{x\in\R^n:|x|_{\setS}\leq\rho(\|w_t\|)\}
\end{flalign}
where $\rho$ is a $K_\infty$-function and $w_t$ is the truncation of $w$ at time $t$ [i.e., $w_t(\tau)=w(\tau)$ if $0\leq \tau\leq t$, and $w_t(\tau)=0$ if $\tau\geq t$].

\begin{definition}[Input-to-State Safety]
System (\ref{eq:iss-system}) is input-to-state safe (ISSf) on $\setS$ if, for any $x_0\in\setS$ and any $w\in L_\infty^m$, there is a $\rho\in K_\infty$ such that $x(t,x_0,w)\in\setS_w$ for all $t\geq0$. Herein, we refer to $\rho$ to as the ISSf gain.
\end{definition}

\begin{remark}
The ISSf of system (\ref{eq:iss-system}) implies that the maximal safety violation over time interval $[0,t)$ is bounded by $\rho(\|w_t\|)$. Note that, in the ISSf literature \cite{kolathaya2018input, lyu2022small, lyu2023highsmall, alan2023control}, the set $\setS_w$ is defined as
\begin{flalign}\label{eq:ISSf-set-old}
    \setS_w=\{x\in\R^n:h(x)+\eta(\|w_t\|)\geq0\}.
\end{flalign}
In fact, (\ref{eq:setS-w-new-def}) and (\ref{eq:ISSf-set-old}) are equivalent. On one hand, with the combination of (\ref{eq:barrier-radial-unbounded}) and (\ref{eq:ISSf-set-old}), $\alpha_1(-|x|_\setS)\geq\min\{h(x),0\}\geq-\eta(\|w_t\|)$, which implies (\ref{eq:setS-w-new-def}) with $\rho(s)=-\alpha_1^{-1}(-\eta(s))$. On the other hand, it follows from (\ref{eq:barrier-radial-unbounded}) and (\ref{eq:setS-w-new-def}) that $h(x)\geq\min\{h(x), 0\}\geq\alpha_2(-|x|_{\setS})\geq\alpha_2(-\rho(\|w_t\|))$, which gives (\ref{eq:ISSf-set-old}) with $\eta(s)=-\alpha_2(-\rho(s))$.
\end{remark}

\begin{definition}[Input-to-State Stability \cite{sontag1995characterizations}]
System (\ref{eq:iss-system}) is input-to-state stable (ISS) with respect to a set $\setS$ if there exist functions $\beta\in KL$ and $\rho\in K_\infty$ such that
\begin{flalign}
    |x(t,x_0,w)|_{\setS}\leq\beta(|x_0|_{\setS},t)+\rho(\|w_t\|),\;\;\;\;\forall x_0\in\R^n,\;\;\forall t\geq0.
\end{flalign}
\end{definition}

\begin{definition}[ISSf Barrier Function\cite{kolathaya2018input, lyu2022small, krstic2023inverse}]\label{def:ISSf-BF}
A continuously differentiable function $h:\R^n\rightarrow\R$ satisfying (\ref{eq:safety-and-bf-amespaper}) is said to be an ISSf barrier function (ISSf-BF) for system (\ref{eq:iss-system}) if, for any $x\in\R^n$, one of the following holds:
\begin{flalign}
    &\nabla h(x)f(x,w)\geq-\alpha(h(x))-\rho(|w|)
    \label{eq:def-disspass-issfbf}\\
    &|h(x)|\geq\rho(|w|)\Rightarrow\nabla h(x)f(x,w)\geq-\alpha(h(x))
    \label{eq:def-lyu-issfbf}\\
    &\min\{0,h(x)\}\leq-\rho(|w|)\Rightarrow\nabla h(x)f(x,w)\geq-\alpha(h(x))
    \label{eq:krstic-GM-cond}
\end{flalign}
where $\alpha\in EK_\infty$ and $\rho\in K_\infty$.
\end{definition}

The ISSf-BF conditions in (\ref{eq:def-disspass-issfbf}), (\ref{eq:def-lyu-issfbf}) and (\ref{eq:krstic-GM-cond}) were given in \cite{kolathaya2018input}, \cite{lyu2022small} and \cite{krstic2023inverse}, respectively, where it is shown that the existence of an ISSf-BF is a sufficient condition for ISSf.  For the case of $w\equiv0$, ISSf-BFs reduce to ZBFs. As will be shown later, such ISSf-BF definitions are equivalent.

\subsection{Inverse Optimal ISSf Gain Assignment}

This subsection revisits the inverse optimal ISSf gain assignment problem \cite{krstic2023inverse}. Consider
\begin{flalign}\label{eq:aff-contr-sys-distur}
    \dot{x}=f(x)+g_1(x)w+g_2(x)u
\end{flalign}
where $x\in\R^n$ is the state, $w\in\R^{m_1}$ is the disturbance input, $u\in\R^{m_2}$ is the control input, and $f:\R^n\rightarrow\R^n$, $g_1:\R^n\rightarrow\R^n\times\R^{m_1}$ and $g_2:\R^n\rightarrow\R^n\times\R^{m_2}$ are locally Lipschitz. Let $u_0$ be the nominal controller used to accomplish the control task (e.g., tracking or stabilization).

\begin{definition}[Inverse Optimal ISSf Gain Assignment \cite{krstic2023inverse}]\label{def:inverse-optimal-ISSf}
By the inverse ISSf gain assignment problem of (\ref{eq:aff-contr-sys-distur}), we mean that there exist
\begin{itemize}
  \item a continuously differentiable $K_\infty$-function $\gamma$ whose derivative $\gamma'$ is also a $K_\infty$-function,
      \item a matrix-value function $R(x,u_0)$ satisfying $R(x,u_0)=R(x,u_0)^\T>0$,
  \item a continuous function $l(x,u_0)$ satisfying $l(x,u_0)>0$ for all $x\in\Int(\setS)$, $l(x,u_0)<0$ for all $x\in\R^n\backslash\setS$, and $l(x,u_0)\rightarrow-\infty$ as $|x|_{\setS}\rightarrow+\infty$,
  \item a safety-critical controller $u=k(x)$
\end{itemize}
such that i) the closed-loop system is ISSf on $\setS$ with a gain $\rho\in K_\infty$, and ii) the following cost functional is maximized:
\begin{flalign}\label{eq:cost-func-barrier}
    &J(u)
    =\inf_{w\in L_\infty^m}\Bigg\{\lim_{t\rightarrow+\infty}\Bigg[4 h(x(t))
        +\int_{0}^{t}\Bigg(l(x,u_0)
     \nonumber\\
    &\;\;\;\;\;\;\;\;\;\;\;\;\;\;\;\;
    -(u-u_0)^TR(x,u_0)(u-u_0)+2\lambda\gamma\Bigg(\frac{|w|}{\lambda}\Bigg)\Bigg)d\tau\Bigg]\Bigg\}
\end{flalign}
where $\lambda>0$ is a constant.
\end{definition}

To handle the disturbance input $w$ and establish inverse optimality, introduce an auxiliary system for (\ref{eq:aff-contr-sys-distur}) as follows
\begin{flalign}\label{eq:affine-auxiliary-system}
    \dot{x}=f(x)-g_1(x)\ell\gamma(2|L_{g_1}h|)\frac{(L_{g_1}h)^\T}{|L_{g_1}h|^2}+g_2(x)u
\end{flalign}
where $\gamma$ is a continuously differentiable $K_\infty$-function selected by control designers, and $\ell\gamma$ denotes the Legendre-Fenchel transform of $\gamma$:
\begin{flalign}
    \ell\gamma(r)=\int_0^r(\gamma')^{-1}(s)ds.
\end{flalign}
The properties of the Legendre-Fenchel transform can be found in Appendix III.

\begin{lemma}[\cite{krstic2023inverse}]\label{lemma:krst-inverse}
Suppose that, for any nominal controller $u_0$, there is a matrix-valued function $R(x,u_0)=R(x,u_0)^\T>0$ such that
\begin{flalign}\label{eq:LGH-ctrl}
    u
    =u_0+\bar{u}(x,u_0)
    :=u_0+ R(x,u_0)^{-1}(L_{g_2}h)^\T
\end{flalign}
is a safe controller for the auxiliary system (\ref{eq:affine-auxiliary-system}) such that
\begin{flalign}\label{eq:stanrd-krstic-ISSf-CBF}
    &L_{f+g_2u_0}h-\ell\gamma(2|L_{g_1}h|)
        \nonumber\\
    &\;\;\;\;\;\;\;\;+L_{g_2}hR(x,u_0)^{-1}(L_{g_2}h)^\T\geq-\alpha(h(x)),
    \;\;\forall x\in\R^n
\end{flalign}
where $\alpha$ is an $EK_\infty$-function. Then
\begin{flalign}\label{eq:krstic-inv-opti-safety-filter}
    u
    &=u_0+\bar{u}^*(x,u_0)
        \nonumber\\
    &:=u_0+2\bar{u}(x,u_0)
        \nonumber\\
    &=u_0+2R(x,u_0)^{-1}(L_{g_2}h)^\T
\end{flalign}
solves the inverse optimal ISSf gain assignment of (\ref{eq:aff-contr-sys-distur}) by maximizing the cost functional (\ref{eq:cost-func-barrier}) with $\lambda\in(0,2]$ and
\begin{flalign}\label{eq:def-of-lxu}
    l(x,u_0)
    &=-4\Big[L_{f+g_2u_0}h-\ell\gamma(2|L_{g_1}h|)
        \nonumber\\
    &\;\;\;\;\;\;\;\;\;\;
        +L_{g_2}hR(x,u_0)^{-1}(L_{g_2}h)^\T\Big]
        \nonumber\\
    &\;\;\;\;\;\;\;\;\;\;
        -2(2-\lambda)\ell\gamma(2|L_{g_1}h|).
\end{flalign}
\end{lemma}

As pointed out in \cite{krstic2023inverse}, many safe controllers can be written in the form of (\ref{eq:LGH-ctrl}), including the famous CBF-QP controller \cite{ames2014control}.

\begin{lemma}[\cite{krstic2023inverse}]\label{lem:cbf-qp-to-Lgh}
The CBF-QP problem
\begin{flalign}\label{eq:ISSf-CBF-QP}
    &\bar{u}_{QP}
    ={\arg\min}_{u\in\R^{m_2}}|u-u_0|^2
        \nonumber\\
    &\text{s.t.}\;\;\;\;L_fh-\ell\gamma(2|L_{g_1}h|)+L_ghu\geq-\alpha(h(x))
\end{flalign}
has an explicit solution
\begin{flalign}\label{eq:bar-u-QP-def}
    u
    =u_0+\bar{u}_{QP}
    :=u_0+R(x,u_0)^{-1}(L_{g_2}h)^2
\end{flalign}
with
\begin{flalign}
    R(x,u_0)
    &=\frac{|L_{g_2}h|^2}{\max\{0,-\omega\}},
        \\
    \omega(x,u_0)
    &=L_{f+g_2u_0}h-\ell\gamma(2|L_{g_1}h|)+\alpha(h(x)).
        \label{eq:CBF-QP-Omega}
\end{flalign}
Moreover, the controller $u=u_0+2\bar{u}_{QP}(x,u_0)$ solves the inverse optimal ISSf gain assignment of (\ref{eq:aff-contr-sys-distur}).
\end{lemma}

\section{Converse ISSf Barrier Function Theorem}\label{sec:ISSf-BF-thm}

This section first reveals the equivalence among different ISSf-BF definitions, and then proposes a converse ISSf-BF theorem to establish the equivalence between ISSf and ISSf-BFs.

\subsection{Equivalence Among ISSf-BFs}

The following lemma establishes the equivalence among ISSf-BF conditions given in (\ref{eq:def-disspass-issfbf}), (\ref{eq:def-lyu-issfbf}) and (\ref{eq:krstic-GM-cond}).

\begin{lemma}\label{eq:equi-ISSf-krstic-ames}
Suppose that the safe set $\setS$ is compact. Then,
\begin{flalign}
    (\ref{eq:def-disspass-issfbf}) \Leftrightarrow (\ref{eq:def-lyu-issfbf}) \Leftrightarrow (\ref{eq:krstic-GM-cond}).
    \nonumber
\end{flalign}
\end{lemma}

\noindent
\textbf{Proof.}
(\ref{eq:def-disspass-issfbf}) $\Leftrightarrow$ (\ref{eq:def-lyu-issfbf}) can be verified with the proof of \cite[Lemma 1]{lyu2023highsmall}. (\ref{eq:def-disspass-issfbf}) $\Rightarrow$ (\ref{eq:krstic-GM-cond}) is straightforward. If the implication (\ref{eq:krstic-GM-cond}) $\Rightarrow$ (\ref{eq:def-disspass-issfbf}) is true, then it follows from the equivalences above that (\ref{eq:def-lyu-issfbf}) $\Leftrightarrow$ (\ref{eq:krstic-GM-cond}).

Now, we prove (\ref{eq:krstic-GM-cond}) $\Rightarrow$ (\ref{eq:def-disspass-issfbf}). Given an ISSf-BF $h(x)$ satisfying (\ref{eq:krstic-GM-cond}), we have the following three cases.

Case I: $h(x)<-\rho(|w|)$. Clearly,
\begin{flalign}\label{eq:ISS-equi-caseone}
    \nabla h(x)f(x,w)
    &\geq-\alpha(h(x))
        \nonumber\\
    &\geq-\alpha(h(x))-\tilde{\rho}_1(|w|)
\end{flalign}
for any arbitrary $\tilde{\rho}_1\in K_\infty$.

Case II: $-\rho(|w|)\leq h(x)<0$. In this case,
\begin{flalign}\label{eq:ISS-equi-casetwo}
    &\nabla h(x)f(x,w)+\alpha(h(x))
     \nonumber\\
    &\;\;\;\;\;\;\;\;
     \geq\nabla h(x)f(x,w)+\alpha(-\rho(|w|))
     \nonumber\\
    &\;\;\;\;\;\;\;\;
     \geq\min\{0,\nabla h(x)f(x,w)\}+\alpha(-\rho(|w|))
     \nonumber\\
    &\;\;\;\;\;\;\;\;
     \geq-\tilde{\rho}_2(|w|)
\end{flalign}
where
\begin{flalign}
    \tilde{\rho}_2(r)=-\inf_{x\in\{x:-\rho(r)\leq h(x)\leq0\},\;|w|<r}\min\{0,\nabla h(x)f(x,w)\}-\alpha(-\rho(r)).
\end{flalign}
Clearly, $\tilde{\rho}_2$ is strictly increasing for all $r\geq0$. By the smoothness of $h(x)$ and (\ref{eq:barrier-radial-unbounded}), we have $\nabla h(x)=0$ for all $x\in\partial\setS$. Thus, $\tilde{\rho}_2(0)=0$. Because $\alpha\in EK_\infty$ and $\rho\in K_\infty$, $\tilde\rho_2(r)\rightarrow+\infty$ as $r\rightarrow+\infty$. In summary, $\tilde\rho_2$ is a $K_\infty$-function

Case III: $h(x)\geq0$. When $w\equiv0$, $\min\{0,h(x)\}=0=-\rho(|w|)$, which, together with (\ref{eq:krstic-GM-cond}), implies
\begin{flalign}
    \nabla h(x)f(x,0)\geq-\alpha(h(x)),\;\;\forall x\in\setS.
\end{flalign}
Thus,
\begin{flalign}
    \nabla h(x)f(x,w)
    &=\nabla h(x)f(x,0)+\nabla h(x)[f(x,w)-f(x,0)]
     \nonumber\\
    &\geq-\alpha(h(x))-|\nabla h(x)|\cdot|f(x,w)-f(x,0)|
     \nonumber\\
    &\geq-\alpha(h(x))-\gamma(h(x),|w|),\;\;\forall x\in\setS
\end{flalign}
where
\begin{flalign}
    \gamma(r,s)=r+s+\sup_{x\in\{x:0\leq h(x)\leq r\},\;|w|\leq s}|\nabla h(x)|\cdot|f(x,w)-f(x,0)|.
\end{flalign}
Clearly, $\gamma(r,0)=0$ and $\gamma(r,s)$ is strictly increasing on each arguments. Additionally, because $\nabla h(x)=0$ when $x\in\partial\setS$, we have $\gamma(0,s)=0$. Thus, $r\mapsto\gamma(r, s)$ and $s\mapsto\gamma(r,s)$ are of class $K$. According to \cite[Corollary IV.5]{angeli2000characterization}, there exists a function $\tilde\gamma\in K$ such that $\gamma(r,s)\leq\tilde\gamma(r)\tilde\gamma(s)$. Therefore,
\begin{flalign}\label{eq:ISS-equi-casethree}
    \nabla h(x)f(x,w)
    &\geq-\alpha(h(x))-\tilde\gamma(h(x))\tilde\gamma(|w|)
     \nonumber\\
    &\geq-\alpha(h(x))-\frac{1}{2}\tilde\gamma(h(x))^2-\frac{1}{2}\tilde\gamma(|w|)^2
     \nonumber\\
    &\geq-\alpha(h(x))-\frac{1}{2}\tilde\gamma(h(x))^2-\tilde\rho_3(|w|),\;\;\forall x\in\setS
\end{flalign}
where $\tilde\rho_3$ is a $K_\infty$-function satisfying $\tilde\rho_3(s)\geq\tilde\gamma(s)^2/2$ for all $s\geq0$. Let
\begin{flalign}
    \tilde\alpha(s)
    &=
    \left\{
      \begin{array}{ll}
        \alpha(s)+\frac{1}{2}\tilde\gamma(s)^2, & s\geq0 \\
        \alpha(s), & s<0
      \end{array}
    \right.
    \\
    \tilde{\rho}(s)&=\max\{\tilde\rho_1(s),\tilde\rho_2(s),\tilde\rho_3(s)\}.
\end{flalign}
It follows from (\ref{eq:ISS-equi-caseone})-(\ref{eq:ISS-equi-casethree}) that
\begin{flalign}
    \nabla h(x)f(x,w)\geq-\tilde\alpha(h(x))-\tilde\rho(|w|),\;\;\forall x\in\R^n
\end{flalign}
which is identical to (\ref{eq:def-disspass-issfbf}), and thus (\ref{eq:krstic-GM-cond}) $\Rightarrow$ (\ref{eq:def-disspass-issfbf}) follows.
\hfill $\Box$
\vskip5pt

\subsection{Converse ISSf-BF Theorem}

The following converse ISSf-BF theorem establishes the equivalence between ISSf and ISSf-BFs.

\begin{theorem}\label{thm:converse-ISSF-BF}
Suppose that system (\ref{eq:iss-system}) is forward complete and there is a compact set $\setS$ such that $\setS\bigcap\X_u=\emptyset$. Then system (\ref{eq:iss-system}) is ISSf on $\setS$ and ISS with respect to $\setS$ if and only if it has an ISSf-BF.
\end{theorem}

Because $w$ is not necessarily bounded, Theorem \ref{thm:converse-ISSF-BF} cannot be proved with the analysis in \cite{prajna2007convex, wisniewski2015converse, ratschan2018converse, liu2021converse, maghenem2022converse, lyu2023converseZBF}. To prove Theorem \ref{thm:converse-ISSF-BF}, we need to introduce two technical lemmas. Denote by $M_D$ the set of signal $d(\cdot)$ such that $|d|\leq1$. Let $\rho$ be the corresponding ISSf gain. Take
\begin{flalign}
    \bar\rho(s)=k\max\{\rho(s),s\},
\end{flalign}
where $k>1$ is a constant. Obviously, $\bar\rho^{-1}$ is locally Lipschitz and satisfies $\rho\circ\bar{\rho}^{-1}(s)<s$. Let
\begin{flalign}
    \varphi(x)=
    \left\{
      \begin{array}{ll}
        \bar\rho^{-1}(|x|_{\R^n\backslash\setS}), & x\in\setS \\
        -\bar\rho^{-1}(|x|_{\setS}), & x\in\R^n\backslash\setS
      \end{array}
    \right.
\end{flalign}
Introduce the auxiliary system
\begin{flalign}\label{eq:iss-aux-system}
    \dot{x}(t)=f(x(t),d(t)\varphi(x(t))):=g(x(t),d(t))
\end{flalign}
with $d\in M_D$. Denote the solution of (\ref{eq:iss-aux-system}) by $x_\varphi(t,x_0,d)$. We say that $\setS$ is a robustly forward invariant set for system (\ref{eq:iss-aux-system}), if for any $x_0\in\setS$ and $d\in M_D$, $x_\varphi(t,x_0,d)\in\setS$ for all $t\geq0$. Moreover, $\setS$ is a uniformly globally asymptotically stable (UGAS) set for (\ref{eq:iss-aux-system}), if there exists $\beta\in KL$ such that
\begin{flalign}
    |x_\varphi(t,x_0,d)|_{\setS}\leq\beta(|x_0|_{\setS},t),\;\;\forall x_0\in\R^n.
\end{flalign}

\begin{lemma}\label{lemma:auxi-sys-safe}
Suppose that system (\ref{eq:iss-system}) is ISSf on $\setS$. Then $\setS$ is a robustly forward invariant and UGAS set for the auxiliary system (\ref{eq:iss-aux-system}).
\end{lemma}

The proof of Lemma \ref{lemma:auxi-sys-safe} is given in Appendix I.

\begin{lemma}\label{converse-robust-ZBF}
If $\setS$ is a robustly forward invariant and UGAS set for the auxiliary system (\ref{eq:iss-aux-system}), then there is a continuously differentiable function $h:\R^n\rightarrow\R$ such that
\begin{flalign}\label{eq:robust-ZBF}
    \nabla h(x)g(x,d)\geq-\alpha(h(x)),\;\;\forall x\in\R^n,\;\;\forall d\in M_D
\end{flalign}
with $\alpha\in EK_\infty$.
\end{lemma}

Lemma \ref{converse-robust-ZBF} is a straightforward extension of \cite[Theorem 1]{lyu2023converseZBF} to the cases involving bounded disturbance inputs. Its proof is given in Appendix II for readability. With the two lemmas above, we are now ready to prove Theorem \ref{thm:converse-ISSF-BF}.

\noindent
\textbf{Proof of Theorem \ref{thm:converse-ISSF-BF}.}
By applying Lemma \ref{converse-robust-ZBF} to (\ref{eq:iss-aux-system}), it follows that
\begin{flalign}
    \nabla h(x)f(x,d\varphi(x))
    \geq-\alpha(h(x)),\;\;\forall x\in\R^n,\;\forall d\in M_D.
\end{flalign}
Hence, for any $w$ satisfying $|w|\leq|\varphi(x)|$,
\begin{flalign}
    \nabla h(x)f(x,w)\geq-\alpha(h(x)),\;\;\forall x\in\R^n.
\end{flalign}
Let $\hat\rho$ be a $K_\infty$-function such that
\begin{flalign}
    \hat\rho^{-1}(s)=\min\{\bar\rho^{-1}(\alpha_2^{-1}(s)),\bar\rho^{-1}(-\alpha_2^{-1}(-s))\}.
\end{flalign}
Recalling (\ref{eq:barrier-radial-unbounded}), we get $\hat\rho^{-1}(|h(x)|)\leq|\varphi(x)|$. Therefore,
\begin{flalign}\label{eq:gain-margin-conv-ISSF-BF}
    |h(x)|
    \geq\hat\rho(|w|)
    \Rightarrow\nabla h(x)f(x,w)\geq-\alpha(h(x)),\;\;\forall x\in\R^n,
\end{flalign}
which is identical to (\ref{eq:def-lyu-issfbf}). Additionally, with Lemma \ref{eq:equi-ISSf-krstic-ames}, it is not difficult to prove the existence of a converse ISSf-BF $h(x)$ satisfying (\ref{eq:def-disspass-issfbf}) or (\ref{eq:krstic-GM-cond}).
\hfill $\Box$
\vskip5pt

\section{Inverse Optimality of ISSf Controllers}

In this section, we employ the converse ISSf-BF theorem established in Section \ref{sec:ISSf-BF-thm} to explore the relationship between the existence of an ISSf controller and the solvability of the inverse optimal ISSf gain assignment problem of system
\begin{flalign}\label{eq:aff-contr-sys-distur-suff-necess-cond}
    \dot{x}=f(x)+g_1(x)w+g_2(x)u.
\end{flalign}

\begin{theorem}\label{thm:inverse-optimal}
System (\ref{eq:aff-contr-sys-distur-suff-necess-cond}) is rendered ISSf on $\setS$ and ISS with respect to $\setS$ by a safety-critical controller if and only if the inverse optimal ISSf gain assignment problem is solvable.
\end{theorem}

\begin{remark}
As can be seen in Lemma \ref{lemma:krst-inverse}, Theorem 4 in \cite{krstic2023inverse} focuses on the connection between ISSf-CBFs and inverse optimality. In contrast, Theorem \ref{thm:inverse-optimal} establishes that the existence of an ISSf controller is both a necessary and sufficient condition for the solvability of the inverse optimal ISSf gain assignment problem of system (\ref{eq:aff-contr-sys-distur-suff-necess-cond}).
\end{remark}

\noindent
\textbf{Proof of Theorem \ref{thm:inverse-optimal}.}
{\it Necessity.}
Because the inverse optimal ISSf gain assignment problem for (\ref{eq:aff-contr-sys-distur-suff-necess-cond}) is solvable, the following HJI equation holds:
\begin{flalign}
    0&=\max_u\min_w\Bigg\{L_fh(x)+L_{g_1}h(x)w+L_{g_2}h(x)u
     \nonumber\\
     &\;\;\;\;\;\;\;\;\;\;\;\;\;\;
      +l(x,u_0)-(u-u_0)^\T R(x,u_0)(u-u_0)+2\lambda\gamma\Bigg(\frac{|w|}{\lambda}\Bigg)\Bigg\}
\end{flalign}
which, together with Lemma \ref{eq:yong-inequality} in Appendix III, implies
\begin{flalign}\label{eq:hji-equation}
    &L_{f+g_2u_0}h+\frac{1}{4}L_{g_2}hR(x,u_0)^{-1}(L_{g_2}h)^\T
     \nonumber\\
    &\;\;\;\;\;\;\;\;\;\;\;\;\;\;\;\;\;\;\;\;\;\;\;\;\;\;\;\;
     -2\lambda\ell\gamma\Bigg(\frac{|L_{g_1}h|}{2}\Bigg)+l(x,u_0)=0.
\end{flalign}
Differentiating $h(x)$ along the solution of the closed-loop system (\ref{eq:aff-contr-sys-distur-suff-necess-cond}) with controller
\begin{flalign}\label{eq:ctrl-inv-opt-necess}
    u=u_0+R(x,u_0)^{-1}(L_{g_2}h)^T,
\end{flalign}
we have
\begin{flalign}
    \dot{h}(x)\Big|_{(\ref{eq:aff-contr-sys-distur-suff-necess-cond}), (\ref{eq:ctrl-inv-opt-necess})}
    &=L_{f+gu_0}h
     +L_{g_1}hw
     +L_{g_2}hR(x,u_0)^{-1}(L_{g_2}h)^\T
      \nonumber\\
    &\geq L_{f+gu_0}h
        +L_{g_2}hR(x,u_0)^{-1}(L_{g_2}h)^\T
      \nonumber\\
    &\;\;\;\;\;\;\;\;\;\;
        -2\lambda\ell{\gamma}(|L_{g_1}h|)
        -2\lambda\gamma\Bigg(\frac{|w|}{\lambda}\Bigg).
\end{flalign}
With Lemma \ref{lemma:property-LF} in Appendix III,
\begin{flalign}
    \dot{h}(x)\Big|_{(\ref{eq:aff-contr-sys-distur-suff-necess-cond}), (\ref{eq:ctrl-inv-opt-necess})}
    &\geq L_{f+gu_0}h
        +L_{g_2}hR(x,u_0)^{-1}(L_{g_2}h)^\T
      \nonumber\\
    &\;\;\;\;\;\;\;\;\;\;
        -2\lambda\ell{\gamma}\Bigg(\frac{|L_{g_1}h|}{2}\Bigg)
        -2\lambda\gamma\Bigg(\frac{|w|}{\lambda}\Bigg).
\end{flalign}
By combining this with (\ref{eq:hji-equation}), we have
\begin{flalign}\label{eq:dh-lx-gamma}
    \dot{h}(x)\Big|_{(\ref{eq:aff-contr-sys-distur-suff-necess-cond}), (\ref{eq:ctrl-inv-opt-necess})}
    \geq-l(x,u_0)-2\lambda\gamma\Bigg(\frac{|w|}{\lambda}\Bigg).
\end{flalign}
Take
\begin{flalign}
    \hat\alpha(r)
    =
    \left\{
      \begin{array}{ll}
        \sup_{0\leq h(x)\leq r}l(x,u_0), & r>0 \\
        0, & r=0 \\
        \sup_{h(x)\leq r}l(x,u_0), & r<0
      \end{array}
    \right.
\end{flalign}
{Because $l(x,u_0)$ is continuous on $\R^n$, positive on $\Int(\setS)$, and negative on $\R^n\backslash\setS$, $\hat\alpha(r)$ is non-decreasing for all $r\in\R$. Moreover, since $l(x,u_0)\rightarrow-\infty$ as $|x|_{\setS}\rightarrow+\infty$, we can pick a function $\hat\alpha$ such that $\hat\alpha(r)\rightarrow-\infty$ as $r\rightarrow-\infty$. With the definition of $\hat\alpha$, we have $\hat\alpha(h(x))\geq l(x,u_0)$. Let $\alpha$ be an $EK$-function such that $\alpha(r)\geq\hat\alpha(r)$ and $\alpha(r)\rightarrow-\infty$ as $r\rightarrow-\infty$. Then, with (\ref{eq:dh-lx-gamma}),
\begin{flalign}
    \dot{h}(x)\Big|_{(\ref{eq:aff-contr-sys-distur-suff-necess-cond}), (\ref{eq:ctrl-inv-opt-necess})}
    \geq-\alpha(h(x))-2\lambda\gamma\Bigg(\frac{|w|}{\lambda}\Bigg)
\end{flalign}
which implies that $h(x)$ is an ISSf-BF for the closed-loop system (\ref{eq:aff-contr-sys-distur-suff-necess-cond}), (\ref{eq:ctrl-inv-opt-necess}). Therefore, there is a safety-critical controller rendering system (\ref{eq:aff-contr-sys-distur}) ISSf on $\setS$ and ISS with respect to $\setS$.

{\it Sufficiency.}
Because system (\ref{eq:aff-contr-sys-distur-suff-necess-cond}) can be rendered ISSf on $\setS$ and ISS with respect to $\setS$, according to Theorem \ref{thm:converse-ISSF-BF}, there is an ISSf-CBF $h(x)$ such that (\ref{eq:barrier-radial-unbounded}) holds for some $EK$-functions $\alpha_1,\alpha_2$ that tend to $-\infty$ as $r\rightarrow-\infty$, and
\begin{flalign}
    \min\{0,h(x)\}
    &\leq-\gamma(|w|)
     \nonumber\\
    &\Rightarrow\sup_{u\in\R^m}\{L_fh(x)+L_{g_1}h(x)w+L_{g_2}h(x)u\}\geq-\alpha(h(x))
\end{flalign}
with $\gamma\in K_\infty$ and $\alpha\in EK_\infty$. Take $V(x)=\max\{0,-h(x)\}$ for convenience. Let
\begin{flalign}
    \omega(x,u_0)
    =L_{f+g_2u_0}h(x)-|L_{g_1}h(x)|\rho^{-1}(V(x))
     +\alpha(h(x)).
\end{flalign}
Consider the following Sontag-type controller as in \cite[Theorem 1]{krstic2023inverse}:
\begin{flalign}\label{eq:sontag-controller}
    {u}_{S}
    =
    \left\{
      \begin{array}{ll}
        (L_{g_2}h(x))^\T\kappa(x,u_0), & (L_{g_2}h(x))^\T\neq0 \\
        0, & (L_{g_2}h(x))^\T=0
      \end{array}
    \right.
\end{flalign}
where
\begin{flalign}
    \kappa(x,u_0)
    =\frac{-\omega+\sqrt{\omega^2+(L_{g_2}h(L_{g_2}h)^\T)^2}}{L_{g_2}h(L_{g_2}h)^\T}.
\end{flalign}

In the following, we are ready to prove that the safety-critical controller
\begin{equation}\label{eq:sontag-ctrl}
    u=u_0+ u_S(x,u_0)
\end{equation}
solves the inverse optimal ISSf gain assignment problem of (\ref{eq:aff-contr-sys-distur}). To this end, we need to show
\begin{itemize}
  \item the cost functional $J(u)$ in (\ref{eq:cost-func-barrier}) is maximized by (\ref{eq:sontag-ctrl});
  \item the closed-loop system (\ref{eq:aff-contr-sys-distur-suff-necess-cond}) with (\ref{eq:sontag-ctrl}) is ISSf on $\setS$ with a gain $\rho\in K_\infty$;
  \item the function $l(x,u_0)$ in (\ref{eq:def-of-lxu}) satisfies the desired properties.
\end{itemize}

\emph{Step 1: Proving that the cost functional $J(u)$ in (\ref{eq:cost-func-barrier}) is maximized.} Take
\begin{flalign}
    R(x,u_0)=\frac{2L_{g_2}h(L_{g_2}h)^\T}{-\omega+\sqrt{\omega^2+(L_{g_2}h(L_{g_2}h)^\T)^2}}.
\end{flalign}
Then the Sontag-type controller $u_S$ in (\ref{eq:sontag-controller}) can be rewritten as $u_S=2R(x,u_0)^{-1}(L_{g_1}h(x))^\T$. Following the derivation of \cite[Eq. (47)]{krstic2023inverse}, it is not difficult to verify that the safety-critical controller (\ref{eq:sontag-ctrl}) maximizes the cost functional $J(u)$ in (\ref{eq:cost-func-barrier}).

\emph{Step 2: Proving that the closed-loop system is ISSf on $\setS$ with a gain $\rho\in K_\infty$.} Differentiating $h(x)$ along the closed-loop system (\ref{eq:aff-contr-sys-distur-suff-necess-cond}), (\ref{eq:sontag-ctrl}), we have
\begin{flalign}
    \dot{h}(x)\Big|_{(\ref{eq:aff-contr-sys-distur-suff-necess-cond}), (\ref{eq:sontag-ctrl})}
    &=L_{f+g_2u_0}h(x)+L_{g_1}h(x)w+L_{g_2}h(x)u_S
     \nonumber\\
    &=L_{f+g_2u_0}h(x)+L_{g_1}h(x)w
     \nonumber\\
    &\;\;\;\;\;\;\;\;\;\;\;\;\;\;
     -\omega+\sqrt{\omega^2+(L_{g_2}h(L_{g_2}h)^\T)^2}
     \nonumber\\
    &\geq
        -\alpha(h(x))
        +|L_{g_1}h|(\rho^{-1}(V)-|w|),
\end{flalign}
and thus,
\begin{flalign}
    \min\{0,h(x)\}\leq-\rho(|w|)\Rightarrow\dot{h}(x)\geq-\alpha(h(x))
\end{flalign}
which, together with Theorem \ref{thm:converse-ISSF-BF}, implies that the closed-loop system (\ref{eq:aff-contr-sys-distur-suff-necess-cond}), (\ref{eq:sontag-ctrl}) is ISSf on $\setS$ and ISS with respect to $\setS$.

\emph{Step 3: Proving that $l(x,u_0)$ in (\ref{eq:def-of-lxu}) is as desired.} Note that
\begin{flalign}\label{eq:Lh-R}
    &L_{f+g_2u_0}h(x)+L_{g_2}hR(x,u_0)^{-1}(L_{g_2}h)^\T
     \nonumber\\
    &\;\;\;\;\;\;\;\;
     =L_{f+g_2u_0}h(x)+\frac{1}{2}\Bigg(-\omega+\sqrt{\omega^2+(L_{g_2}h(L_{g_2}h)^\T)^2}\Bigg)
     \nonumber\\
    &\;\;\;\;\;\;\;\;
     =\frac{1}{2}L_{f+g_2u_0}h(x)+\frac{1}{2}|L_{g_1}h(x)|\rho^{-1}(V(x))
      \nonumber\\
    &\;\;\;\;\;\;\;\;\;\;\;\;
     -\frac{1}{2}\alpha(h(x))+\frac{1}{2}\sqrt{\omega^2+(L_{g_2}h(L_{g_2}h)^\T)^2}
      \nonumber\\
    &\;\;\;\;\;\;\;\;
     =-\alpha(h(x))
     +|L_{g_1}h(x)|\rho^{-1}(V(x))
     \nonumber\\
    &\;\;\;\;\;\;\;\;\;\;\;\;
      +\frac{1}{2}\underbrace{\Big(L_{f+gu_0}h(x)-|L_{g_1}h(x)|\rho^{-1}(V(x))+\alpha(h(x))\Big)}_{\omega(x,u_0)}
      \nonumber\\
    &\;\;\;\;\;\;\;\;\;\;\;\;
      +\frac{1}{2}\sqrt{\omega^2+(L_{g_2}h(L_{g_2}h)^\T)^2}
      \nonumber\\
    &\;\;\;\;\;\;\;\;
     =
      -\alpha(h(x))
      +|L_{g_1}h(x)|\rho^{-1}(V(x))
      \nonumber\\
    &\;\;\;\;\;\;\;\;\;\;\;\;
      +\frac{1}{2}\Bigg(\omega+\sqrt{\omega^2+(L_{g_2}h(L_{g_2}h)^\T)^2}\Bigg)
      \nonumber\\
    &\;\;\;\;\;\;\;\;
    \geq
      -\alpha(h(x))
      +|L_{g_1}h(x)|\rho^{-1}(V(x)).
\end{flalign}
Then we have the following two cases.

Case I: $x\in\setS$. We have $V(x)=0$. From (\ref{eq:Lh-R}),
\begin{flalign}\label{eq:lxu0-bound-prelimin-A}
    L_{f+g_2u_0}h+L_{g_2}hR(x,u_0)^{-1}(L_{g_2}h)^\T
    \geq
      -\alpha(h(x)).
\end{flalign}
Thus,
\begin{flalign}
    &L_{f+g_2u_0}h-\ell\gamma(2|L_{g_1}h|)+L_{g_2}hR(x,u_0)^{-1}(L_{g_2}h)^\T
     \nonumber\\
    &\;\;\;\;\;\;\;\;\;\;\;\;\;\;\;\;\;\;\;\;\;\;\;\;\;\;\;\;\;\;\;\;\;\;\;\;\;\;\;\;
     \geq -\alpha(h(x))-\ell\gamma(2|L_{g_1}h|).
\end{flalign}

Case II: $x\in\R^n\backslash\setS$. In this case, $h(x)=-V(x)$. Because $|L_{g_1}h(x)|$ vanishes at $x\in\partial\setS$, there exists a $\pi\in EK_\infty$ such that
\begin{flalign}
    |L_{g_1}h(x)|
    \leq\pi(|x|_{\setS})
    \leq\pi(-\alpha_1^{-1}(-V(x)))
    :=\hat\pi(V(x)),\;\;\forall x\in\R^n\backslash\setS
    \nonumber
\end{flalign}
where $\alpha_1$ was given in (\ref{eq:barrier-radial-unbounded}). Let
\begin{flalign}
    \varrho(r)=\int_{0}^{r/2}\rho^{-1}\circ\hat\pi^{-1}(s)ds.
\end{flalign}
Clearly, $\varrho$ is a $K_\infty$ function and satisfies $\varrho(r)\leq r\rho^{-1}\circ\hat\pi^{-1}(r/2)$. Let $\gamma=\ell\varrho$. By Lemma \ref{lemma:property-LF}, $\ell\ell\varrho=\varrho$, and thus,
\begin{flalign}
    \ell\gamma(2r)=\varrho(2r)\leq r\rho^{-1}\circ\hat\pi^{-1}(r).
\end{flalign}
Combining this with (\ref{eq:Lh-R}) yields
\begin{flalign}\label{eq:lxu0-bound-prelimin-B}
    &L_{f+g_2u_0}h-\ell\gamma(2|L_{g_1}h|)+L_{g_2}hR(x,u_0)^{-1}(L_{g_2}h)^\T
     \nonumber\\
    &\;\;\;\;
     \geq L_{f+g_2u_0}h
        -|L_{g_1}h|\rho^{-1}\circ\hat\pi^{-1}(|L_{g_1}h|)
        +L_{g_2}hR(x,u_0)^{-1}(L_{g_2}h)^\T
    &\;\;\;\;
     \nonumber\\
    &\;\;\;\;
     \geq L_{f+g_2u_0}h-|L_{g_1}h|\rho^{-1}(V(x))+L_{g_2}hR(x,u_0)^{-1}(L_{g_2}h)^\T
     \nonumber\\
    &\;\;\;\;
     \geq-\alpha(h(x)).
\end{flalign}
Take
\begin{flalign}
    \mu(r)=
    \left\{
      \begin{array}{ll}
        \alpha(r)+\sup_{0\leq h(x)\leq r}\ell\gamma(2|L_{g_1}h|), & r\geq0 \\
        \alpha(r), & r<0
      \end{array}
    \right.
\end{flalign}
Because $\sup_{0\leq h(x)\leq r}\ell\gamma(2|L_{g_1}h|)$ is non-decreasing, $\mu$ is an $EK$-function satisfying $\mu(r)\rightarrow-\infty$ as $r\rightarrow-\infty$. By combining (\ref{eq:def-of-lxu}), (\ref{eq:lxu0-bound-prelimin-A}) and (\ref{eq:lxu0-bound-prelimin-B}), we have
\begin{flalign}
    l(x,u_0)
    &\leq-4\Big[L_{f+g_2u_0}h(x)-\ell\gamma(2|L_{g_1}h(x)|)
     \nonumber\\
    &\;\;\;\;\;\;\;\;\;\;\;\;\;\;\;\;\;\;\;\;\;\;
     +L_{g_2}h(x)R(x,u_0)^{-1}(L_{g_2}h(x))^\T\Big]
     \nonumber\\
    &\leq-4\mu(h(x))
\end{flalign}
which implies that $l(x,u_0)$ is the function as desired.
\hfill $\Box$
\vskip5pt

\section{Gain Margins of Inverse Optimal Safe Controllers under Zero Disturbance}

The remainder of this paper investigates the robustness of inverse optimal ISSf controllers with respect to variations in the control gain. To clearly explain to the readers why such controllers possess a gain margin, this section considers the case of zero disturbance. We first discuss the gain margin of inverse optimal safe controllers derived from standard CBF conditions \cite{krstic2023inverse, ames2016control}, and then propose a method to further improve the gain margin property. This improvement is the key idea toward guaranteeing the gain margin of $[1/2,\infty)$ for inverse optimal ISSf controllers in the next section.

Consider the zero-disturbance affine control system
\begin{flalign}\label{eq:contr-sys-zeroDistur-GM}
    \dot{x}=f(x)+g_2(x)u.
\end{flalign}
We say that a safe controller for system (\ref{eq:contr-sys-zeroDistur-GM}) has gain margin $[\sigma_1,\sigma_2)$ if, under the same controller, the system
\begin{flalign}\label{eq:contr-sys-zeroDistur-GM-sigma}
    \dot{x}=f(x)+\sigma g_2(x)u
\end{flalign}
remains safe for any uncertain control gain $\sigma\in[\sigma_1,\sigma_2)$. Let $h(x)$ be a continuously differentiable CBF candidate of (\ref{eq:contr-sys-zeroDistur-GM-sigma}). For convenience, we say that the open-loop vector field $f(x)$ acts safely (resp. unsafely) on the safety boundary $\partial\setS$, if $L_fh>0$ (resp. $L_fh<0$) for all $x\in\partial\setS$. Similarly, we say that the nominal controller $u_0$ acts safely (resp. unsafely) on $\partial\setS$, if $L_{g_2}hu_0>0$ (resp. $L_{g_2}hu_0<0$) for all $x\in\partial\setS$.

\subsection{Gain Margin of Standard Inverse Optimal Safe Controllers}

The result below reveals the gain margin property of the standard inverse optimal safe controller proposed in \cite{krstic2023inverse}.

\begin{theorem}\label{thm:krstic-gain-margin}
Let $h(x)$ be a CBF candidate and $u_0$ be the nominal controller. Suppose that there is a matrix-valued function $R(x,u_0)=R(x,u_0)^\T>0$ such that
\begin{flalign}\label{eq:krstic-inv-opti-sf-ctrl-GM}
    u
    =u_0+\bar{u}^*(x,u_0)
    =u_0+2R(x,u_0)^{-1}(L_{g_2}h)^\T
\end{flalign}
is an inverse optimal safe controller for system (\ref{eq:contr-sys-zeroDistur-GM}) in the sense of maximizing the cost functional
\begin{flalign}\label{eq:cost-zero-disturbance}
    J(u)
    &=\lim_{t\rightarrow+\infty}\Bigg[4 h(x(t))
        \nonumber\\
    &\;\;\;\;
    +\int_{0}^{t}\Big(l(x,u_0)-(u-u_0)^TR(x,u_0)(u-u_0)\Big)d\tau\Bigg]
\end{flalign}
with
\begin{flalign}\label{eq:def-of-lxu-zero-input}
    l(x,u_0)
    =-4\Big[L_{f+g_2u_0}h
        +L_{g_2}hR(x,u_0)^{-1}(L_{g_2}h)^\T\Big]\leq4\alpha(h(x)),
\end{flalign}
where $\alpha$ is an $EK_\infty$ function. Then we have:
\begin{basedescript}{\desclabelstyle{\pushlabel}\desclabelwidth{0.7cm}}
\item[\hspace{0.17cm}(i)] For the case where both $f(x)$ and $u_0$ act unsafely on the safety boundary $\partial\setS$, the inverse optimal safe controller (\ref{eq:krstic-inv-opti-sf-ctrl-GM}) has gain margin $[1/2,\infty)$. Moreover, if $\sigma\in[1/2,\infty)$, the inverse optimal system (\ref{eq:contr-sys-zeroDistur-GM-sigma})-(\ref{eq:krstic-inv-opti-sf-ctrl-GM}) is LAS with respect to $\setS$.
\item[\hspace{0.17cm}(ii)] For the case where $f(x)$ acts unsafely but $u_0$ acts safely on $\partial\setS$, the inverse optimal safe controller (\ref{eq:krstic-inv-opti-sf-ctrl-GM}) has gain margin $[1,\infty)$. Moreover, if $\sigma\in[1,\infty)$, the inverse optimal system (\ref{eq:contr-sys-zeroDistur-GM-sigma})-(\ref{eq:krstic-inv-opti-sf-ctrl-GM}) is LAS with respect to $\setS$.
\item[\hspace{0.17cm}(iii)] For the case where $f(x)$ acts safely but $u_0$ acts unsafely on $\partial\setS$, the inverse optimal safe controller (\ref{eq:krstic-inv-opti-sf-ctrl-GM}) has gain margin $[1/2,1]$. Moreover, if $\sigma\in[1/2,1]$, the inverse optimal control system (\ref{eq:contr-sys-zeroDistur-GM-sigma})-(\ref{eq:krstic-inv-opti-sf-ctrl-GM}) is LAS with respect to $\setS$.
\end{basedescript}
\end{theorem}

\noindent
\textbf{Proof.}
Since (\ref{eq:krstic-inv-opti-sf-ctrl-GM}) is an inverse optimal safe controller for system (\ref{eq:contr-sys-zeroDistur-GM}) with respect to cost functional (\ref{eq:cost-func-barrier}), the following HJI equation holds:
\begin{flalign}\label{eq:HJB-Krstic}
    L_{f+g_2u_0}h+L_{g_2}hR(x,u_0)^{-1}(L_{g_2}h)^\T+\frac{l(x,u_0)}{4}=0.
\end{flalign}
Differentiating $h(x)$ along the solution of closed-loop system (\ref{eq:contr-sys-zeroDistur-GM-sigma}), (\ref{eq:krstic-inv-opti-sf-ctrl-GM}) with uncertain control gain $\sigma$, we have
\begin{flalign}\label{eq:diff-krstic-HJB-solution}
    \dot{h}(x)\Big|_{(\ref{eq:contr-sys-zeroDistur-GM-sigma}), (\ref{eq:krstic-inv-opti-sf-ctrl-GM})}
    =L_fh+\sigma L_{g_2}h\Big(u_0+2R(x,u_0)^{-1}(L_{g_2}h)^\T\Big),\;\;\forall x\in\R^n.
\end{flalign}
Examine the following three cases.

(i) {\it Both $f(x)$ and $u_0$ act unsafely on $\partial\setS$.} By combining (\ref{eq:HJB-Krstic}) and (\ref{eq:diff-krstic-HJB-solution}), we have
\begin{flalign}\label{eq:dh-Thm3-fUnsafe-u0Unsafe-A}
    \dot{h}(x)\Big|_{(\ref{eq:contr-sys-zeroDistur-GM-sigma}), (\ref{eq:krstic-inv-opti-sf-ctrl-GM})}
    =-\frac{\sigma}{2}l(x,u_0)
        -(2\sigma-1)L_fh
        -\sigma L_{g_2}hu_0,\;\;\forall x\in\R^n.
\end{flalign}
Let $\delta>0$ be a constant that $L_fh(x)<0$ and $L_{g_2}h(x)u_0<0$ always hold on the set
\begin{flalign}\label{eq:attr-reg-proof-thm-inv-opt}
    R_A^\delta = \{x\in\R^n:-\delta<h(x)\leq0\}.
\end{flalign}
Take
\begin{flalign}\label{eq:ROA}
    D=\setS\cup R_A^\delta.
\end{flalign}
Let $\hat\alpha_1$ and $\hat\alpha_2$ be $EK$ functions such that
\begin{flalign}
    \hat\alpha_1(h(x))
    &\geq L_{f}h(x),\;\;\;\;\;\forall x\in D,
        \label{eq:dh-Thm3-fUnsafe-u0Unsafe-hatAlpha1}\\
    \hat\alpha_2(h(x))
    &\geq L_{g_2}h(x)u_0,\;\;\forall x\in D.
        \label{eq:dh-Thm3-fUnsafe-u0Unsafe-hatAlpha2}
\end{flalign}
Because $L_fh(x)<0$ and $L_{g_2}h(x)u_0<0$ on $\partial\setS$, such $EK$ functions always exist. If $\sigma\in[1/2,\infty)$, substituting (\ref{eq:dh-Thm3-fUnsafe-u0Unsafe-hatAlpha1})-(\ref{eq:dh-Thm3-fUnsafe-u0Unsafe-hatAlpha2}) into (\ref{eq:dh-Thm3-fUnsafe-u0Unsafe-A}) yields
\begin{flalign}
    &\dot{h}(x)\Big|_{(\ref{eq:contr-sys-zeroDistur-GM-sigma}), (\ref{eq:krstic-inv-opti-sf-ctrl-GM})}
        \nonumber\\
    &\;\;
     \geq-\frac{\sigma}{2}l(x,u_0)
        -(2\sigma-1)\hat\alpha_1(h(x))
        -\sigma \hat\alpha_2(h(x))
        \nonumber\\
    &\;\;
     \geq-2\sigma\alpha(h(x))
        -(2\sigma-1)\hat\alpha_1(h(x))
        -\sigma \hat\alpha_2(h(x)),
        \;\forall x\in D,
\end{flalign}
which, together with Lemma \ref{lemma:zbf-safety}, implies that, if $\sigma\in[1/2,\infty)$, the inverse optimal system (\ref{eq:contr-sys-zeroDistur-GM-sigma}), (\ref{eq:krstic-inv-opti-sf-ctrl-GM}) is safe on $\setS$, and LAS with respect to $\setS$ where the region of attraction is $D$.

(ii) {\it $f(x)$ acts unsafely but $u_0$ acts safely on $\partial\setS$.} By combining (\ref{eq:HJB-Krstic}) and (\ref{eq:diff-krstic-HJB-solution}), we have
\begin{flalign}
    \dot{h}(x)\Big|_{(\ref{eq:contr-sys-zeroDistur-GM-sigma}), (\ref{eq:krstic-inv-opti-sf-ctrl-GM})}
    &=-\frac{l(x,u_0)}{4}+(\sigma-1)L_{g_2}hu_0
        \nonumber\\
    &\;\;\;\;\;\;\;\;\;
        +(2\sigma-1) L_{g_2}hR(x,u_0)^{-1}(L_{g_2}h)^\T,\;\;\;\;\forall x\in\R^n.
        \label{eq:dh-Thm3-fUnsafe-u0Safe-A}
\end{flalign}
Let $\delta>0$ be a constant such that $L_{g_2}h(x)u_0>0$ on the set $R_A^\delta$ defined in (\ref{eq:attr-reg-proof-thm-inv-opt}). Because $L_{g_2}h(x)u_0>0$ for all $x\in\partial\setS$, there exists an $EK$ function $\hat\alpha_2$ such that
\begin{flalign}
    L_{g_2}h(x)u_0
    \geq -\hat\alpha_2(h(x)),\;\;\;\;\;\forall x\in D
        \label{eq:dh-Thm3-fUnsafe-u0Safe-hatAlpha2}
\end{flalign}
where $D$ was defined in (\ref{eq:ROA}). Then, by substituting (\ref{eq:dh-Thm3-fUnsafe-u0Safe-hatAlpha2}) into (\ref{eq:dh-Thm3-fUnsafe-u0Safe-A}) yields that, if $\sigma\in[1,\infty)$,
\begin{flalign}
    \dot{h}(x)\Big|_{(\ref{eq:contr-sys-zeroDistur-GM-sigma}), (\ref{eq:krstic-inv-opti-sf-ctrl-GM})}
    &\geq-\frac{l(x,u_0)}{4}
        -(\sigma-1)\hat\alpha_2(h(x))
        \nonumber\\
    &\geq-\alpha(h(x))-(\sigma-1)\hat\alpha_2(h(x)),\;\;\;\;\forall x\in D
\end{flalign}
which, together with Lemma \ref{lemma:zbf-safety} implies that, if $\sigma\in[1,\infty)$, the inverse optimal system (\ref{eq:contr-sys-zeroDistur-GM-sigma}), (\ref{eq:krstic-inv-opti-sf-ctrl-GM}) is safe on $\setS$, and LAS with respect to $\setS$ where the region of attraction is $D$.

(iii) {\it $f(x)$ acts safely while $u_0$ acts unsafely on $\partial\setS$.} Rearrange (\ref{eq:diff-krstic-HJB-solution}) as
\begin{flalign}
    \dot{h}(x)\Big|_{(\ref{eq:contr-sys-zeroDistur-GM-sigma}), (\ref{eq:krstic-inv-opti-sf-ctrl-GM})}
    &=L_{f+g_2u_0}h
        -(1-\sigma)L_{g_2}hu_0
        \nonumber\\
    &\;\;\;\;\;\;\;\;
        +2\sigma L_{g_2}hR(x,u_0)^{-1}(L_{g_2}h)^\T,
        \;\;\;\;\forall x\in\R^n.
        \label{eq:dh-Thm3-fSafe-u0Unsafe-A}
\end{flalign}
By combining (\ref{eq:HJB-Krstic}) and (\ref{eq:dh-Thm3-fSafe-u0Unsafe-A}), we have
\begin{flalign}
    \dot{h}(x)\Big|_{(\ref{eq:contr-sys-zeroDistur-GM-sigma}), (\ref{eq:krstic-inv-opti-sf-ctrl-GM})}
    &=-\frac{l(x,u_0)}{4}
        -(1-\sigma)L_{g_2}hu_0
        \nonumber\\
    &\;\;\;\;\;\;\;\;\;\;\;\;\;\;
        +(2\sigma-1)L_{g_2}hR(x,u_0)^{-1}(L_{g_2}h)^\T
        \nonumber\\
    &\geq-\alpha(h(x))
        -(1-\sigma)L_{g_2}h
        \nonumber\\
    &\;\;\;\;\;\;\;\;\;\;\;\;\;\;
        \times\Big(u_0+R(x,u_0)^{-1}(L_{g_2}h)^\T\Big),
        \;\;\;\;
        \forall x\in\R^n.
        \label{eq:dh-Thm3-fSafe-u0Unsafe-B}
\end{flalign}
Because $L_fh<0$ on $\partial\setS$, it follows from (\ref{eq:HJB-Krstic}) that
\begin{flalign}
        L_{g_2}h\Big(u_0+R(x,u_0)^{-1}(L_{g_2}h)^\T\Big)<0,\;\;\;\;\forall x\in\partial\setS.
\end{flalign}
Hence, there exists an $EK$ function $\hat\alpha_3$ such that
\begin{flalign}
        -L_{g_2}h\Big(u_0+R(x,u_0)^{-1}(L_{g_2}h)^\T\Big)\geq-\hat\alpha_3(h(x)),
        \;\;\;\;
        \forall x\in D
\end{flalign}
where $D$ was defined in (\ref{eq:ROA}). Consequently, when $\sigma\in[1/2,1]$,
\begin{flalign}
    \dot{h}(x)\Big|_{(\ref{eq:contr-sys-zeroDistur-GM-sigma}), (\ref{eq:krstic-inv-opti-sf-ctrl-GM})}
    \geq-\alpha(h(x))
        -(1-\sigma)\hat\alpha_3(h(x)),
        \;\;\;\;
        \forall x\in D
\end{flalign}
which, together with Lemma \ref{lemma:zbf-safety}, implies that, if $\sigma\in[1/2,1]$, the inverse optimal system (\ref{eq:contr-sys-zeroDistur-GM-sigma})-(\ref{eq:krstic-inv-opti-sf-ctrl-GM}) is safe on $\setS$, and LAS with respect to $\setS$ where the region of attraction is $D$.
\hfill $\Box$
\vskip5pt

\begin{remark}\label{rem:essential-GM-standard-inv-opt}
We now provide some intuition for Theorem \ref{thm:krstic-gain-margin}.
\begin{itemize}
  \item When $f(x)$ acts safely on the safety boundary $\partial\setS$, the joint action of $f(x)$ and the inverse optimal safety override $\bar{u}^*$ can counteract the unsafe effect of the nominal controller $u_0$. Reducing $\sigma$ weakens both the unsafe effect of $u_0$ and the safe effect of $\bar{u}^*$. Since the influence of $f(x)$ remains unchanged, this reduction acts to amplify the combined effect of $f(x)$ and $\bar{u}^*$. Hence, it is reasonable for the closed-loop system (\ref{eq:contr-sys-zeroDistur-GM-sigma})-(\ref{eq:krstic-inv-opti-sf-ctrl-GM}) to remain safe if $\sigma$ changes within $[1/2,1]$.
  \item When $f(x)$ acts unsafely but $u_0$ acts safely on $\partial\setS$, increasing $\sigma$ enhances the safe effect of the overall control $u_0+\bar{u}^*$. Consequently, the closed-loop system remains safe for all $\sigma \in [1,\infty)$.
  \item When both $f(x)$ and $u_0$ act unsafely on $\partial\setS$, the override part $\bar{u}^*$ has an excess of safety to dominate the unsafe effect of $u_0$. Hence, the overall control $u_0+\bar{u}^*$ can be increased arbitrary since, as the unsafe action of $u_0$ is increased, the action of $\bar{u}^*$ is increased even more. The inverse optimal safe controller $u_0+\bar{u}^*$ contains twice the safety override needed to counteract the unsafe effects of $f(x)$ and $u_0$. Hence, reducing its gain by up to half still retains the required override to maintain safety, while simultaneously mitigating the unsafe influence of $u_0$. In summary, the resulting closed-loop system can maintain safety whenever $\sigma$ changes within $[1/2,\infty)$.
\end{itemize}
\end{remark}

\begin{remark}
In Theorem \ref{thm:krstic-gain-margin}, the LAS property of $\setS$ implies that the closed-loop safety is sensitive to large additive disturbances. If a sufficiently large (but bounded) disturbance drives the state trajectory outside the region of attraction, the resulting safety violation may grow unbounded.
\end{remark}

As pointed out in Remark \ref{rem:essential-GM-standard-inv-opt}, the standard inverse optimal safe controller $u=u_0+\bar{u}^*=u_0+2\bar{u}$  has twice the safety override of the naive CBF-QP controller ${u}=u_0+\bar{u}_{QP}=u_0+\bar{u}$. By combining this observation with Theorem \ref{thm:krstic-gain-margin}, we obtain the following corollary.

\begin{corollary}\label{cor:CBF-GM}
For the naive CBF-QP controller
\begin{flalign}\label{eq:GM-CBF-QP}
    u=u_0+\bar{u}_{QP}
\end{flalign}
with
\begin{flalign}
    \bar{u}_{QP}
    &={\arg\min}_{u\in\R^{m_2}}|u-u_0|^2
        \nonumber\\
    &\text{s.t.}\;\;\;\;L_fh+L_{g_2}hu\geq-\alpha(h(x)),
\end{flalign}
the following properties hold:
\begin{basedescript}{\desclabelstyle{\pushlabel}\desclabelwidth{0.7cm}}
\item[\hspace{0.17cm}(i)] For the case where $f(x)$ acts unsafely on the safety boundary $\partial\setS$, the CBF-QP controller (\ref{eq:GM-CBF-QP}) has gain margin $[1,\infty)$. Moreover, if $\sigma\in[1,\infty)$, the closed-loop system (\ref{eq:contr-sys-zeroDistur-GM-sigma}) with (\ref{eq:GM-CBF-QP}) is LAS with respect to $\setS$.
\item[\hspace{0.17cm}(ii)] For the case where $f(x)$ acts safely while $u_0$ is unsafe on $\partial\setS$, the CBF-QP controller (\ref{eq:GM-CBF-QP}) has gain margin $[1/2,1]$. Moreover, if $\sigma\in[1/2,1]$, the closed-loop system (\ref{eq:contr-sys-zeroDistur-GM-sigma}) with (\ref{eq:GM-CBF-QP}) is LAS with respect to $\setS$.
\end{basedescript}
\end{corollary}

\noindent
\textbf{Proof.}
By Lemma \ref{lem:cbf-qp-to-Lgh}, the CBF-QP controller (\ref{eq:GM-CBF-QP}) can be written as
\begin{flalign}\label{eq:CBF-control-rewrite}
    u=u_0+R(x,u_0)^{-1}(L_{g_2}h)^\T.
\end{flalign}
(i) \emph{$f(x)$ acts unsafely on $\partial\setS$}. Differentiating $h(x)$ along the closed-loop system (\ref{eq:contr-sys-zeroDistur-GM-sigma}), (\ref{eq:CBF-control-rewrite}),
\begin{flalign}
    \dot{h}(x)\Big|_{(\ref{eq:contr-sys-zeroDistur-GM-sigma}), (\ref{eq:CBF-control-rewrite})}
    &=L_fh+\sigma L_{g_2}h\Big(u_0+R(x,u_0)^{-1}(L_{g_2}h)^\T\Big)
        \label{eq:CBF-ctrl-GM-dh-prop1-A}
\end{flalign}
Using (\ref{eq:HJB-Krstic}), we have
\begin{flalign}
    \dot{h}(x)\Big|_{(\ref{eq:contr-sys-zeroDistur-GM-sigma}), (\ref{eq:CBF-control-rewrite})}
    =-\frac{\sigma}{4}l(x,u_0)
        -(\sigma-1)L_fh,\;\;\;\;\forall x\in\R^n.
\end{flalign}
Similar to the proof of Theorem \ref{thm:krstic-gain-margin}(i), we can find an open set $D$ containing $\setS$ and a function $\hat\alpha\in EK$ such that $\hat\alpha_1(h(x))\geq L_fh(x)$ for all $x\in D$. Consequently,
\begin{flalign}
    \dot{h}(x)\Big|_{(\ref{eq:contr-sys-zeroDistur-GM-sigma}), (\ref{eq:CBF-control-rewrite})}
    =-\sigma\alpha(h(x))
        -(\sigma-1)\hat\alpha_1(h(x)),\;\;\;\;\forall x\in D
\end{flalign}
which establishes Corollary \ref{cor:CBF-GM}(i).

(ii) \emph{$f(x)$ acts safely but $u_0$ acts unsafely on $\partial\setS$}. By substituting (\ref{eq:HJB-Krstic}) into (\ref{eq:CBF-ctrl-GM-dh-prop1-A}), we have
\begin{flalign}
    \dot{h}(x)\Big|_{(\ref{eq:contr-sys-zeroDistur-GM-sigma}), (\ref{eq:CBF-control-rewrite})}
    &=-\frac{l(x,u_0)}{4}-(1-\sigma)L_{g_2}h
        \nonumber\\
    &\;\;\;\;\;\;\;\;\;\;
        \times
        \Big(u_0+R(x,u_0)^{-1}(L_{g_2}h)^\T\Big)
        \nonumber\\
    &\geq-\alpha(h(x))
        -(1-\sigma)L_{g_2}h
        \nonumber\\
    &\;\;\;\;\;\;\;\;\;\;
        \times
        \Big(u_0+R(x,u_0)^{-1}(L_{g_2}h)^\T\Big),
        \;\;\;
        \forall x\in\R^n.
        \label{eq:CBF-ctrl-GM-dh-prop2}
\end{flalign}
Then Corollary \ref{thm:krstic-gain-margin}(ii) follows by an argument similar
to that used in the proof of Theorem \ref{thm:krstic-gain-margin}(i).
\hfill $\Box$
\vskip5pt

In the following, we are ready to show that the constraints on $\sigma$ in Theorem \ref{thm:krstic-gain-margin} cannot be relaxed. The following example shows that an inverse optimal CBF-QP controller may be unable to tolerate any reduction in the control gain when $f(x)$ acts unsafely but $u_0$ acts safely on $\partial\setS$.

\begin{example}[$f(x)$ acts unsafely but $u_0$ acts safely on $\partial\setS$]\label{example:cannot-gain-reduction}
Consider the system
\begin{flalign}\label{eq:exmample-1-system}
    \dot{x}=-11.1x-1+ u.
\end{flalign}
The safety constraint is $x\geq0$ and the norminal control task is to make $x(t)\rightarrow 0$. Design the nominal controller as
\begin{flalign}\label{eq:uo-of-fUnsafe-u0Safe-example}
    u_0=10.1x+1
\end{flalign}
Select the CBF candidate $h(x)=x$. It follows that $L_fh<0$ and $L_ghu_0>0$ when $x\in\partial\setS$, which implies that the open-loop vector field $f(x)$ acts unsafely on the boundary $\partial\setS$, while the nominal controller $u_0$ acts safely. With Lemma \ref{lem:cbf-qp-to-Lgh}, we can compute that $\omega=0$ and the CBF-QP controller
\begin{flalign}\label{eq:exm-1-CBF-QP}
    u
    =u_0+2\max\{0,-\omega\}
    =10.1x+1
\end{flalign}
is inverse optimal for system (\ref{eq:exmample-1-system}) by maximizing the cost functional (\ref{eq:cost-func-barrier}) with $l(x,u_0)=4x$, $R(x,u_0)=+\infty$, and $\gamma(s)=0$.
When there are uncertainties in the control gain, (\ref{eq:exmample-1-system}) becomes
\begin{flalign}\label{eq:exm-1-system-uncertain}
    \dot{x}=-11.1x-1+ \sigma u.
\end{flalign}
Substituting (\ref{eq:exm-1-CBF-QP}) into (\ref{eq:exm-1-system-uncertain}), we see that the closed-loop system
\begin{flalign}\label{eq:exmp1-inv-opt-closeloop}
    \dot{x}=(10.1\sigma-11.1)x+\sigma-1
\end{flalign}
is safe on $\setS=\{x:x\geq0\}$ when $\sigma\geq1$, but unsafe for all $0<\sigma<1$. Therefore, the CBF-QP controller (\ref{eq:exm-1-CBF-QP}) cannot tolerate any reduction in the control gain, although it is inverse optimal. Additionally, for all $\sigma>11.1/10.1\approx1.1$, the inverse optimal system (\ref{eq:exmp1-inv-opt-closeloop}) is LAS with respect to $\setS$ where the region of attraction is $D=\Big\{x:x>-\frac{\sigma-1}{10.1\sigma-11.1}\Big\}$.
\end{example}

Next, we show that an inverse optimal CBF-QP controller may be unable to tolerate any increase in the control gain, if $f(x)$ acts safely but $u_0$ acts unsafely on $\partial\setS$.

\begin{example}[$f(x)$ acts safely but $u_0$ acts unsafely on $\partial\setS$]\label{example:cannot-gain-increase}
Consider the system
\begin{flalign}\label{eq:exm-2-system}
    \dot{x}=9.1x+1+ u
\end{flalign}
The safety constraint is $x\geq0$ and the nominal control task is to make $x(t)\rightarrow 0$. Design the nominal controller
\begin{flalign}\label{eq:u0-of-fSafe-u0Unsafe-example}
    u_0=-10.1x-1.
\end{flalign}
For the CBF candidate $h(x)=x$, it follows that $L_fh>0$ and $L_ghu_0<0$ when $x\in\partial\setS$. Therefore, $f(x)$ acts safely on $\partial\setS$, while $u_0$ acts unsafely. With Lemma \ref{lem:cbf-qp-to-Lgh}, we obtain that $\omega=0$, and the CBF-QP controller
\begin{flalign}\label{eq:exm-2-CBF-QP-controller}
    u
    =u_0+2\max\{0,-\omega\}
    =-10.1x-1
\end{flalign}
is inverse optimal for system (\ref{eq:exm-2-system}) by maximizing the cost functional (\ref{eq:cost-func-barrier}) with $l(x,u_0)=4x$, $R(x,u_0)=+\infty$, and $\gamma(s)=0$.
When there are uncertainties in the control gain, (\ref{eq:exm-2-system}) becomes
\begin{flalign}\label{eq:exm-2-system-uncertain}
    \dot{x}=9.1x+1+ \sigma u
\end{flalign}
By combining (\ref{eq:exm-2-CBF-QP-controller}) and (\ref{eq:exm-2-system-uncertain}), we see that the closed-loop system
\begin{flalign}\label{eq:exm-2-system-uncertain-closeloop}
    \dot{x}=(9.1-10.1\sigma)x+1-\sigma
\end{flalign}
is safe on $\setS=\{x:x\geq0\}$ when $\sigma\in[1/2,1]$, but is unsafe for all $\sigma>1$. Therefore, the CBF-QP controller (\ref{eq:exm-2-CBF-QP-controller}) cannot tolerate any increase in the control gain, though it is inverse optimal. Additionally, if $\sigma<9.1/10.1\approx0.9$, the inverse optimal system (\ref{eq:exm-2-system-uncertain-closeloop}) is LAS with respect to $\setS$ where the region of attraction is $D=\Big\{x:x>-\frac{1-\sigma}{9.1-10.1\sigma}\Big\}$
\end{example}

\subsection{Gain Margin Improvement}

This subsection aims at improving the gain margin of inverse optimal safe controllers.

\begin{theorem}\label{thm:GM-inv-opt-ctrl-dom-zero-dist}
Still under the conditions in Theorem \ref{thm:krstic-gain-margin}, for the controller
\begin{flalign}\label{eq:dominat-optimal-controller}
    u=u_0+\tilde{\bar{u}}^*(x,u_0)=u_0+2\tilde{R}(x,u_0)^{-1}(L_{g_2}h)^\T
\end{flalign}
with
\begin{flalign}\label{eq:tilde-R}
    \tilde{R}(x,u_0)^{-1}
    =R(x,u_0)^{-1}+\frac{\max\{L_fh,0\}+\max\{L_{g_2}hu_0,0\}}{L_{g_2}h(L_{g_2}h)^\T},
\end{flalign}
we have:
\begin{basedescript}{\desclabelstyle{\pushlabel}\desclabelwidth{0.7cm}}
\item[\hspace{0.17cm}(i)] Controller (\ref{eq:dominat-optimal-controller}) is inverse optimal for system (\ref{eq:contr-sys-zeroDistur-GM}) by maximizing the cost functional
\begin{flalign}\label{eq:new-cost-fun-zero-dist}
    &\tilde{J}(u)
    =\inf_{d\in D}\Bigg\{\lim_{t\rightarrow+\infty}\Bigg[4 h(x(t))
        +\int_{0}^{t}\Big(\tilde{l}(x,u_0)
     \nonumber\\
    &\;\;\;\;\;\;\;\;\;\;\;\;\;\;\;\;\;\;\;\;\;\;\;\;\;\;\;\;\;\;\;\;
    -(u-u_0)^T\tilde{R}(x,u_0)(u-u_0)\Big)d\tau\Bigg]\Bigg\}
\end{flalign}
    where
\begin{flalign}\label{eq:tilde-lxu0-zero-dist}
    \tilde{l}(x,u_0)
    &=-4\Big[L_{f+g_2u_0}h
        \nonumber\\
    &\;\;\;\;\;\;\;\;\;\;\;\;\;\;
        +L_{g_2}h\tilde{R}(x,u_0)^{-1}(L_{g_2}h)^\T\Big].
\end{flalign}
\item[\hspace{0.17cm}(ii)] The safe controller (\ref{eq:dominat-optimal-controller}) processes gain margin $[1/2,\infty)$. Moreover, if $\sigma\in[1/2,\infty)$, the closed-loop system (\ref{eq:contr-sys-zeroDistur-GM}), (\ref{eq:dominat-optimal-controller}) is GAS with respect to $\setS$.
\end{basedescript}
\end{theorem}

\noindent
\textbf{Proof.}
(i) Differentiating $h(x)$ along the solution of closed-loop system (\ref{eq:contr-sys-zeroDistur-GM}) with
\begin{flalign}\label{eq:dominat-CBF-controller}
    u=u_0+\tilde{\bar{u}}^*(x,u_0)=u_0+\tilde{R}(x,u_0)^{-1}(L_{g_2}h)^\T
\end{flalign}
yields
\begin{flalign}
    \dot{h}(x)\Big|_{(\ref{eq:contr-sys-zeroDistur-GM}), (\ref{eq:dominat-CBF-controller})}
    &=L_{f+g_2u_0}h
        +L_{g_2}h\tilde{R}(x,u_0)^{-1}(L_{g_2}h)^{\T}
        \nonumber\\
    &=L_{f+g_2u_0}h
        +L_{g_2}hR(x,u_0)^{-1}(L_{g_2}h)^{\T}
        \nonumber\\
    &\;\;\;\;\;\;\;\;\;\;
        +\max\{L_fh,0\}
        +\max\{L_{g_2}hu_0,0\}
        \nonumber\\
    &\geq-\alpha(h(x)),\;\;\;\;\forall x\in\R^n
\end{flalign}
which, together with Lemma \ref{lemma:krst-inverse} (by taking $\gamma=0$), implies that controller (\ref{eq:dominat-optimal-controller}) is inverse optimal for system (\ref{eq:contr-sys-zeroDistur-GM}) by maximizing cost functional (\ref{eq:new-cost-fun-zero-dist}).

(ii) Since controller (\ref{eq:dominat-optimal-controller}) is inverse optimal for system (\ref{eq:contr-sys-zeroDistur-GM}) with respect to cost functional (\ref{eq:new-cost-fun-zero-dist}), the following HJI-equation is solvable:
\begin{flalign}\label{eq:HJI-redesign-zero-dist}
    L_{f+g_2u_0}h
    +L_{g_2}h\tilde{R}(x,u_0)^{-1}(L_{g_2}h)^\T
    +\frac{\tilde{l}(x,u_0)}{4}=0.
\end{flalign}
By combining (\ref{eq:def-of-lxu}) and (\ref{eq:tilde-R}), we have
\begin{flalign}\label{eq:lxu-and-Lf-Lg-zero-dist}
    \tilde{l}(x,u_0)
    &=l(x,u_0)
    -4\max\{L_fh,0\}
        \nonumber\\
    &\;\;\;\;\;\;\;\;
    -4\max\{L_{g_2}hu_0,0\}
        \nonumber\\
    &\leq4\alpha(h(x))
\end{flalign}
Differentiating $h(x)$ along the solution of closed-loop system (\ref{eq:contr-sys-zeroDistur-GM-sigma}), (\ref{eq:dominat-optimal-controller}), and then, using (\ref{eq:HJI-redesign-zero-dist}) and (\ref{eq:lxu-and-Lf-Lg-zero-dist}), it follows that, for all $\sigma\in[1/2,\infty)$,
\begin{flalign}\label{eq:zero-dist-dh-tilde-override}
    &\dot{h}(x)\Big|_{(\ref{eq:contr-sys-zeroDistur-GM}), (\ref{eq:dominat-optimal-controller})}
        \nonumber\\
    &\;\;\;\;
        =L_fh
        +\sigma L_{g_2}h\Big(u_0+2\tilde{R}(x,u_0)^{-1}(L_{g_2}h)^\T\Big)
        \nonumber\\
    &\;\;\;\;
        =-\frac{\sigma}{2}\tilde{l}(x,u_0)
        +(1-2\sigma)L_fh
        -\sigma L_{g_2}hu_0
        \nonumber\\
    &\;\;\;\;
        \geq-\frac{\sigma}{2}\tilde{l}(x,u_0)
            +(1-2\sigma)\max\{L_fh,0\}
            -\sigma\max\{L_{g_2}hu_0,0\}
        \nonumber\\
    &\;\;\;\;
        \geq-\frac{\sigma}{2}l(x,u_0)
        +\max\{L_fh,0\}
        +\sigma\max\{L_{g_2}hu_0,0\}
        \nonumber\\
    &\;\;\;\;
        \geq-2\sigma\alpha(h(x)),
        \;\;\;\;\forall x\in\R^n.
\end{flalign}
By Lemma \ref{lemma:zbf-safety}, if $\sigma\in[1/2,\infty)$, the inverse optimal system (\ref{eq:contr-sys-zeroDistur-GM-sigma}), (\ref{eq:dominat-optimal-controller}) is safe on $\setS$, and GAS with respect to $\setS$.
\hfill $\Box$
\vskip5pt

\begin{remark}
Compared to Theorem \ref{thm:krstic-gain-margin}, Theorem \ref{thm:GM-inv-opt-ctrl-dom-zero-dist} improves the gain margin property in two aspects: i) the inverse optimal safe controller (\ref{eq:dominat-optimal-controller}) has a $[1/2,\infty)$ gain margin without prior knowledge of whether $f(x)$ or $u_0$ acts safely on $\partial\setS$; and ii) the safe set $\setS$ of the closed-loop system is GAS, rather than LAS. As will be shown in next section, such gain margin improvement is the key step toward ensuring gain margin for inverse optimal ISSf controllers. However, from (\ref{eq:tilde-R}), we see that $\tilde{R}(x,u_0)$ is more conservative than $R(x,u_0)$ of Theorem \ref{thm:krstic-gain-margin}, which implies that the gain margin improvement achieved in Theorem \ref{thm:GM-inv-opt-ctrl-dom-zero-dist} comes at the expense of an increased control effort.
\end{remark}

\begin{remark}
For the case where both $f(x)$ and $u_0$ act unsafely on $\partial\setS$, one can find an $EK$ function $\hat\alpha$ such that
\begin{flalign}
    \max\{L_fh,0\}+\max\{L_{g_2}hu_0,0\}\leq\hat\alpha(h(x)),\;\;\;\;\forall x\in\setS.
\end{flalign}
Let $\delta>0$ be a constant such that $L_fh(x)<0$ and $L_{g_2}h(x)u_0<0$ always hold on the set $R_A^\delta = \{x\in\R^n:-\delta<h(x)\leq0\}$. From (\ref{eq:tilde-R}),
\begin{flalign}
    &R(x,u_0)^{-1}\geq
     \tilde{R}(x,u_0)^{-1}-\frac{\hat\alpha(h(x))}{L_{g_2}h(L_{g_2}h)^\T},\;\;\;\;\forall x\in\setS,
        \\
    &R(x,u_0)^{-1}
    = \tilde{R}(x,u_0)^{-1},
        \;\;\;\;\;\;\;\;\;\;\;\;\;\;\;\;\;\;\;\;\;\;\;\;\;\;\forall x\in R_A^\delta.
\end{flalign}
Because $\hat\alpha(h(x))=0$ for all $x\in\partial\setS$, the term $-\frac{\hat\alpha(h(x))}{L_{g_2}h(L_{g_2}h)^\T}$ does not jeopardize the closed-loop safety. Consequently, when both $f(x)$ and $u_0$ act unsafely on $\partial\setS$, the standard inverse optimal safe controller $u=u_0+\bar{u}^*$ in (\ref{eq:krstic-inv-opti-sf-ctrl-GM}) is safe, provided that the controller $u=u_0+\tilde{\bar{u}}^*$ in (\ref{eq:dominat-CBF-controller}) is safe. Hence, Theorem \ref{thm:krstic-gain-margin}(i) is actually a corollary of Theorem \ref{thm:GM-inv-opt-ctrl-dom-zero-dist}.
\end{remark}

Below, we revisit Example \ref{example:cannot-gain-reduction} and Example \ref{example:cannot-gain-increase}, and show that the inverse optimal ISSf controllers in Theorem \ref{thm:GM-inv-opt-ctrl-dom-zero-dist} can inherit the standard $[1/2,\infty)$ gain margin without requiring prior
knowledge of whether $f(x)$ or $u_0$ acts safely on the boundary, while simultaneously guaranteeing GAS of the safe set.

\begin{example}[Example \ref{example:cannot-gain-reduction} revisited]
Recalling (\ref{eq:uo-of-fUnsafe-u0Safe-example}), we have $u_0=10.1x+1$.
By taking $h(x)=x$ and $\alpha(s)=s$, it then follows from Theorem \ref{thm:GM-inv-opt-ctrl-dom-zero-dist} that
\begin{flalign}
    u
    &=u_0+2\tilde{R}(x,u_0)^{-1}(L_{g_2}h)^\T
        \nonumber\\
    &=\max\{8.1x+1,-12.1x-1,30.3x+3,0\}
        \label{eq:CBF-example-deresign}
\end{flalign}
is an inverse optimal controller for (\ref{eq:aff-contr-sys-distur}) in the sense of maximizing cost functional (\ref{eq:new-cost-fun-zero-dist}). Substituting (\ref{eq:CBF-example-deresign}) into (\ref{eq:exm-1-system-uncertain}), we see that, for all $\sigma\in[1/2,\infty)$, the resulting closed-loop system
\begin{flalign}
    \dot{x}
    &=\max\{(8.1\sigma-11.1)x+\sigma-1,
        \nonumber\\
    &\;\;\;\;\;\;\;\;\;\;\;\;
        -(12.1\sigma+11.1)x-(\sigma+1),
        \nonumber\\
    &\;\;\;\;\;\;\;\;\;\;\;\;
        (30.3\sigma-11.1)x+3\sigma-1,
        \nonumber\\
    &\;\;\;\;\;\;\;\;\;\;\;\;
        -11.1x-1
        \}
        \nonumber\\
    &\geq
        (30.3\sigma-11.1)x+3\sigma-1
\end{flalign}
is safe on $\setS=\{x:x>0\}$, and GAS with respect to $\setS$.
\end{example}

\begin{example}[Example \ref{example:cannot-gain-increase} revisited]
Recalling (\ref{eq:u0-of-fSafe-u0Unsafe-example}), we have $u_0=-10.1x-1$. By taking $h(x)=x$ and $\alpha(s)=s$, it then follows from Theorem \ref{thm:GM-inv-opt-ctrl-dom-zero-dist}, the controller
\begin{flalign}
    &u=u_0+2\tilde{R}(x,u_0)^{-1}(L_{g_2}h)^\T
        =\max\{-3x-2,-x+2\}
        \label{eq:domin-examp-fSafe-u0Unsafe}
\end{flalign}
is inverse optimal with respect to (\ref{eq:new-cost-fun-zero-dist}).
Substituting (\ref{eq:domin-examp-fSafe-u0Unsafe}) into (\ref{eq:exm-2-system-uncertain}) yields that, for all $\sigma\in[1/2,\infty)$, the resulting closed-loop system
\begin{flalign}
    \dot{x}
    &=\max\{(9.1-12.1\sigma)x+1-\sigma,
        \nonumber\\
    &\;\;\;\;\;\;\;\;\;\;\;\;
        (9.1+8.1\sigma)x+1+\sigma,
        \nonumber\\
    &\;\;\;\;\;\;\;\;\;\;\;\;
        (9.1-30.3\sigma)x+1-3\sigma,
        \nonumber\\
    &\;\;\;\;\;\;\;\;\;\;\;\;
        9.1x+1
        \}
\end{flalign}
is safe on $\setS=\{x:x\geq0\}$, and GAS with respect to $\setS$.
\end{example}

\section{Gain Margins of Inverse Optimal ISSf Controllers}

This section turns back to the affine control system with disturbance
\begin{flalign}\label{eq:aff-contr-sys-distur-GM}
    \dot{x}=f(x)+g_1(x)w+g_2(x)u.
\end{flalign}
An ISSf controller for system (\ref{eq:aff-contr-sys-distur-GM}) is said to have a $[\sigma_1,\sigma_2)$ gain margin if, under the same control law, the system
\begin{flalign}\label{eq:affine-system-gain-uncertain}
    \dot{x}=f(x)+g_1(x)w+\sigma g_2(x) u
\end{flalign}
remains ISSf for any uncertain control gain $\sigma\in [\sigma_1,\sigma_2)$. The following result is built on Theorem \ref{thm:GM-inv-opt-ctrl-dom-zero-dist}.

\begin{theorem}\label{thm:GM-domination}
Let $h(x)$ be a CBF candidate and $u_0$ be the nominal controller. Consider the auxiliary system of (\ref{eq:aff-contr-sys-distur-GM})
\begin{flalign}\label{eq:affine-aux-sys-GM-domination}
    \dot{x}=f(x)-g_1(x)\ell_\gamma(2|L_{g_1}h|)\frac{(L_{g_1}h)^\T}{|L_{g_1}h|^2}+g_2(x)u
\end{flalign}
where $\gamma$ is a continuously differentiable $K_\infty$-function selected by control designers. Suppose that there is a matrix-valved function $R(x,u_0)=R(x,u_0)^\T>0$ such that
\begin{flalign}\label{eq:ISSf-CBF-dimonate}
    &L_{f+g_2u_0}h-\ell\gamma(2|L_{g_1}h|)
        \nonumber\\
    &\;\;\;\;\;\;\;\;+L_{g_2}hR(x,u_0)^{-1}(L_{g_2}h)^\T\geq-\alpha(h(x)),
    \;\;\forall x\in\R^n.
\end{flalign}
Then, for the controller
\begin{flalign}\label{eq:dominat-optimal-ISSf-controller}
    u=u_0+2\tilde{R}(x,u_0)^{-1}(L_{g_2}h)^\T
\end{flalign}
with
\begin{equation}\label{eq:tilde-R-ISSf}
    \tilde{R}(x,u_0)^{-1}
    =R(x,u_0)^{-1}+\frac{\max\{L_fh,0\}+\max\{L_{g_2}hu_0,0\}}{L_{g_2}h(L_{g_2}h)^\T},
\end{equation}
the following properties hold:
\begin{basedescript}{\desclabelstyle{\pushlabel}\desclabelwidth{0.7cm}}
\item[\hspace{0.17cm}(i)] Controller (\ref{eq:dominat-optimal-ISSf-controller}) solves the inverse optimal ISSf gain assignment problem of system (\ref{eq:aff-contr-sys-distur-GM}) by maximizing the cost functional
\begin{flalign}\label{eq:new-cost-fun}
    &\tilde{J}(u)
    =\inf_{d\in D}\Bigg\{\lim_{t\rightarrow+\infty}\Bigg[4 h(x(t))
        +\int_{0}^{t}\Bigg(\tilde{l}(x,u_0)
     \nonumber\\
    &\;\;\;\;
    -(u-u_0)^T\tilde{R}(x,u_0)(u-u_0)+2\lambda\gamma\Bigg(\frac{|w|}{\lambda}\Bigg)\Bigg)d\tau\Bigg]\Bigg\}
\end{flalign}
where
\begin{flalign}\label{eq:tilde-lxu0}
    \tilde{l}(x,u_0)
    &=-4\Big[L_{f+g_2u_0}h
        -\ell\gamma(2|L_{g_1}h|)
        \nonumber\\
    &\;\;\;\;\;\;\;\;\;\;\;\;\;\;
        +L_{g_2}h\tilde{R}(x,u_0)^{-1}(L_{g_2}h)^T\Big]
        \nonumber\\
    &\;\;\;\;\;\;\;\;\;\;\;\;\;\;
        -2(2-\lambda)\ell\gamma(2|L_{g_1}h|).
\end{flalign}
\item[\hspace{0.17cm}(ii)] The inverse optimal ISSf controller (\ref{eq:dominat-optimal-ISSf-controller}) has gain margin $[1/2,\infty)$. Moreover, if $\sigma\in[1/2,\infty)$, the inverse optimal system (\ref{eq:affine-system-gain-uncertain}), (\ref{eq:dominat-optimal-ISSf-controller}) is ISS with respect to $\setS$.
\end{basedescript}
\end{theorem}

\noindent
\textbf{Proof.}
Property (i) can be established with the same argument as in the proof of Theorem \ref{thm:GM-inv-opt-ctrl-dom-zero-dist}(i). The rest is to prove property (ii). With (\ref{eq:def-of-lxu}) and (\ref{eq:tilde-lxu0}), we have
\begin{flalign}\label{eq:lxu-and-Lf-Lg-zero-dist-ISSf}
    \tilde{l}(x,u_0)
    &=l(x,u_0)
    -4\max\{L_fh,0\}
        \nonumber\\
    &\;\;\;\;\;\;\;\;
    -4\max\{L_{g_2}hu_0,0\}
        \nonumber\\
    &\leq4\alpha(h(x))
\end{flalign}
Since (\ref{eq:dominat-optimal-controller}) is inverse optimal for system (\ref{eq:aff-contr-sys-distur-GM}) with respect to cost functional (\ref{eq:new-cost-fun}), the following HJI-equation is solvable:
\begin{flalign}\label{eq:HJI-redesign}
    L_{f+g_2u_0}h
    +L_{g_2}h\tilde{R}(x,u_0)^{-1}(L_{g_2}h)^\T-\ell\gamma(2|L_{g_1}h|)+\frac{\tilde{l}(x,u_0)}{4}=0
\end{flalign}
Differentiating $h(x)$ along the solution of closed-loop system (\ref{eq:affine-system-gain-uncertain}), (\ref{eq:dominat-optimal-ISSf-controller}), and then, using (\ref{eq:lxu-and-Lf-Lg-zero-dist-ISSf}) and (\ref{eq:HJI-redesign}), we have
\begin{flalign}
    \dot{h}(x)
    &=L_fh+L_{g_1}hw+\sigma L_{g_2}h\Big(u_0+2\tilde{R}(x,u_0)^{-1}(L_{g_2}h)^\T\Big)
        \nonumber\\
    &=(1-2\sigma)L_fh-\sigma L_{g_2}hu_0
        \nonumber\\
    &\;\;\;\;\;\;\;\;
        -\frac{\sigma}{2}\tilde{l}(x,u_0)
        +L_{g_1}hw
        +2\sigma\ell\gamma(2|L_{g_1}h|)
        \nonumber\\
    &\geq(1-2\sigma)L_fh-\sigma L_{g_2}hu_0
        -\frac{\sigma}{2}\tilde{l}(x,u_0)
        -\gamma\Bigg(\frac{|w|}{2}\Bigg)
        \nonumber\\
    &\geq(1-2\sigma)\max\{L_fh,0\}
            -\sigma\max\{L_{g_2}hu_0,0\}
        \nonumber\\
    &\;\;\;\;\;\;\;\;\;\;
        -\frac{\sigma}{2}\tilde{l}(x,u_0)
        -\gamma\Bigg(\frac{|w|}{2}\Bigg)
        \nonumber\\
    &\geq-2\sigma\alpha(h(x))
        +\max\{L_fh,0\}
        +\sigma\max\{L_{g_2}hu_0,0\}
        -\gamma\Bigg(\frac{|w|}{2}\Bigg)
        \nonumber\\
    &\geq-2\sigma\alpha(h(x))
        -\gamma\Bigg(\frac{|w|}{2}\Bigg),\;\;\forall x\in\R^n.
\end{flalign}
With Theorem \ref{thm:inverse-optimal}, we obtain that, if $\sigma\in[1/2,\infty)$, the closed-loop system (\ref{eq:affine-system-gain-uncertain}), (\ref{eq:dominat-optimal-ISSf-controller}) is ISSf on $\setS$, and ISS with respect to $\setS$.
\hfill $\Box$
\vskip5pt

\section{Conclusion}

This paper studies the inverse optimality and gain margin of ISSf controllers. To reveal the connection between ISS and inverse optimality, we first establish a converse ISSf-BF theorem to show that, for general nonlinear systems, the existence of an ISSf-BF is equivalent to ISSf. With this theorem, we find that the achievability of ISSf by feedback is equivalent to the solvability of a HJI equation associated with a meaningful inverse optimal ISSf gain assignment problem. For gain margin, we find that, in the absence of disturbances, the standard inverse optimal safe controllers have a certain degree of gain margin; however, this gain margin property depends on whether the open-loop vector field or the nominal controller acts safely on the boundary, and the resulting closed-loop systems is LAS with respect to the safe set. To this end, we propose a gain margin improvement approach at the expense of an increased control effort. This improvement allows inverse optimal safe controllers to inherit the standard gain margin of $[1/2,\infty)$, while ensuring GAS of the safe set. With this idea, we further show how inverse optimal ISSf controllers can be improved to inherit the gain margin of $[1/2,\infty)$.

\section*{Appendix I: Proof of Lemma \ref{lemma:auxi-sys-safe}}
\setcounter{subsection}{0}

We first show that $\setS$ is a robustly forward invariant set for auxiliary system (\ref{eq:iss-aux-system}) by contradiction. Assume that this is not true. Denote by $t_1$ the first time instant that $x_\varphi(t,x_0,d)$ leaves $\setS$, namely,
\begin{flalign}
    t_1=\inf\{t\geq0:|x_\varphi(t,x_0,d)|_\setS>0\},\;\;\forall x_0\in \setS.
\end{flalign}
Let $x_0\in\partial \setS$. Because system (\ref{eq:iss-system}) is ISSf on $\setS$,
\begin{flalign}
    \sup_{\tau\in[0,t_1]}|x_\varphi(\tau,x_0,d)|_{\setS}
    &
     \leq\rho(\sup_{\tau\in[0,t]}|d(\tau)\bar{\rho}^{-1}(|x_\varphi(\tau,x_0,d)|_\setS)|)
     \nonumber\\
    &
     \leq\rho\circ\bar{\rho}^{-1}(\sup_{\tau\in[0,t_1]}|x_\varphi(\tau,x_0,d)|_\setS)
        \nonumber\\
    &<\sup_{\tau\in[0,t_1]}|x_\varphi(\tau,x_0,d)|_\setS,
     \;\;\forall 0\leq t\leq t_1.
     \label{eq:x-varphi-bd-ieq-one}
\end{flalign}
which yields a contradiction. Hence, $\setS$ is a robustly forward invariant set of the auxiliary system (\ref{eq:iss-aux-system}).

Next, we prove that $\setS$ is a UGAS set for the auxiliary system (\ref{eq:iss-aux-system}). Because system (\ref{eq:iss-system}) is ISS with respect to $\setS$ with a gain $\rho$, there exist functions $\beta\in KL$ and $\rho \in K$ such that the solution of (\ref{eq:iss-aux-system}) satisfies
\begin{flalign}
    &|x_\varphi(t,x_0,d)|_\setS
        \nonumber\\
    &\;\;\;\;
        \leq\beta(|x_0|_\setS,t-t_0)
        +\rho(\sup_{\tau\in[t_0,+\infty)}|d(\tau)\bar{\rho}^{-1}(|x_\varphi(\tau,x_0,d)|_\setS)|),
     \nonumber\\
    &\;\;\;\;
        \leq\beta(|x_0|_\setS,t-t_0)
        +\rho\circ\bar{\rho}^{-1}(\sup_{\tau\in[t_0,+\infty)}|x_\varphi(\tau,x_0,d)|_\setS),
     \nonumber\\
    &\;\;\;\;\;\;\;\;\;\;
        \forall x_0\in\R^n,\;\forall t\geq t_0\geq0.
\end{flalign}
For any $c\in(0,1)$, let $t_0=c t$. Then
\begin{flalign}\label{def-robust-ISS}
    |x_\varphi(t,x_0,d)|_\setS
    &\leq\beta(|x_0|_\setS,(1-c)t)
     \nonumber\\
    &\;\;\;\;
     +\rho\circ\bar{\rho}^{-1}(\sup_{\tau\in[c t,+\infty)}|x_\varphi(\tau,x_0,d)|_\setS),
     \nonumber\\
    &\;\;\;\;
     \forall x_0\in\R^n,\;\forall t\geq t_0\geq0.
\end{flalign}
Noting $\rho\circ\bar{\rho}^{-1}(s)<s$ and applying \cite[Lemma A.1]{Jiang1994Small} to (\ref{def-robust-ISS}), there exists $\hat\beta\in KL$ such that
\begin{equation}
    |x_\varphi(t,x_0,d)|_\setS\leq\hat\beta(|x_0|_\setS,t),\;\;\forall x_0\in\R^n,\;\forall d\in M_D,\;\forall t\geq0
\end{equation}
which implies that system (\ref{eq:iss-aux-system}) is UGAS with respect to $\setS$.

\section*{Appendix II: Proof of Lemma \ref{converse-robust-ZBF}}
\setcounter{subsection}{0}

Let $\phi(\tau,x,d)$ be the solution of auxiliary system (\ref{eq:iss-aux-system}) passing through $x\in\R^n$ at $\tau=0$. For any $x\in\Int(\setS)$, we have the following two cases:
\begin{itemize}
  \item infinite-time reachability: there is no finite $t\geq0$ such that $\phi(t,x,d)\in\partial\setS$;
  \item finite-time reachability: there is a finite $t=T^*(x)>0$ such that $\phi(t,x,d)\in\partial\setS$.
\end{itemize}

\emph{Case I: Infinite-Time Reachability.} It follows from \cite[Lemma 5]{lyu2023converseZBF} that there exist $\chi_1,\chi_2\in K_\infty$ and a constant $c\geq0$ such that
\begin{flalign}\label{eq:bd-of-solution}
    \frac{1}{|\phi(t,x,d)|_{\cl(\R^n\backslash\setS)}}
    \leq\chi_1(t)+\chi_2\Bigg(\frac{1}{|x|_{\cl(\R^n\backslash\setS)}}\Bigg)+c,\;\;\forall x\in\Int(\setS).
\end{flalign}
Take
\begin{flalign}
    W(t,x)=\inf_{-t\leq\tau\leq0,\;\;d\in M_D}\frac{1}{|\phi(\tau,x,d)|_{\cl(\R^n\backslash\setS)}},\;\;\forall x\in\Int(\setS)
\end{flalign}
Clearly,
\begin{flalign}\label{eq:upper-bd-w-t-xi}
    0\leq W(t,x)\leq \frac{1}{|x|_{\cl(\R^n\backslash\setS)}}.
\end{flalign}
By the definition of $W$, there exist $\tau\in[-t,0]$ and $d\in M_D$ such that
\begin{flalign}
    \frac{1}{|\phi(\tau,x,d)|_{\cl(\R^n\backslash\setS)}}=W(t,x).
\end{flalign}
Let $\tilde{d}$ be a signal such that $\tilde{d}(s)=d(s+\tau)$ for all $s\in\R$. Then, by combining this with (\ref{eq:bd-of-solution}), we have
\begin{flalign}
    \frac{1}{|x|_{\cl(\R^n\backslash\setS)}}
    &=\frac{1}{|\phi(-\tau,\phi(\tau,x,d),\tilde{d})|_{\cl(\R^n\backslash\setS)}}
     \nonumber\\
    &\leq\chi_1(-\tau)+\chi_2\Bigg(\frac{1}{|\phi(\tau,x,d)|_{\cl(\R^n\backslash\setS)}}\Bigg)+c
     \nonumber\\
    &\leq\chi_1(t)+\chi_2(W(t,x))+c,
\end{flalign}
and thus,
\begin{flalign}\label{eq:lower-bd-w-t-xi}
    \chi_1^{-1}\Bigg(\frac{1}{2|x|_{\cl(\R^n\backslash\setS)}}\Bigg)
    \leq
    t+\chi_1^{-1}(W(t,x)+c).
\end{flalign}
Take
\begin{flalign}\label{eq:def-tildeW}
    \tilde{W}(x)=\inf_{t\geq0}\alpha(W(t,x))\e^{t},\;\;\forall x\in\Int(\setS).
\end{flalign}
Therefore,
\begin{flalign}\label{eq:bd-of-tilde-W}
    \eta\Bigg(\frac{1}{|x|_{\cl(\R^n)}}\Bigg)
    \leq\tilde{W}(x)
    \leq\alpha\Bigg(\frac{1}{|x|_{\cl(\R^n\backslash\setS)}}\Bigg)
\end{flalign}
where
\begin{flalign}
    \eta(s)=\exp\Bigg[\frac{\chi_1^{-1}(s/2)}{2}\Bigg],\;\;\alpha(s)=\exp\Bigg[\frac{\chi_1^{-1}(\chi_2(s)+c)}{2}\Bigg].
\end{flalign}
Also, the derivative of $\tilde{W}$ along the solution of (\ref{eq:iss-aux-system}) satisfies
\begin{flalign}\label{eq:LgTildeW}
    &\nabla\tilde{W}(x)g(x,d)
     \nonumber\\
    &\;\;\;\;
        =\limsup_{\varepsilon\rightarrow0^+}\frac{\tilde{W}(\phi(\varepsilon,x,d))-\tilde{W}(x)}{\varepsilon}
     \nonumber\\
    &\;\;\;\;
        =\limsup_{\varepsilon\rightarrow0^+}\frac{1}{\varepsilon}\bigg\{\inf_{t\geq0}\alpha(W(t,\phi(\varepsilon,x,d)))\e^{t}-\inf_{t\geq0}\alpha(W(t,x))\e^{t}\bigg\}
     \nonumber\\
    &\;\;\;\;
        \leq\limsup_{\varepsilon\rightarrow0^+}\frac{1}{\varepsilon}\bigg\{\inf_{t\geq\varepsilon}\alpha(W(t,\phi(\varepsilon,x,d)))\e^{t}-\inf_{t\geq0}\alpha(W(t,x))\e^{t}\bigg\}
     \nonumber\\
    &\;\;\;\;
        =\limsup_{\varepsilon\rightarrow0^+}\frac{1}{\varepsilon}\bigg\{\inf_{t\geq0}\alpha(W(t+\varepsilon,\phi(\varepsilon,x,d)))\e^{t+\varepsilon}-\inf_{t\geq0}\alpha(W(t,x))\e^{t}\bigg\}
     \nonumber\\
    &\;\;\;\;
        \leq\limsup_{\varepsilon\rightarrow0^+}\frac{1}{\varepsilon}\bigg\{\inf_{t\geq0}\alpha(W(t,x))\e^{t+\varepsilon}-\inf_{t\geq0}\alpha(W(t,x))\e^{t}\bigg\}
     \nonumber\\
    &\;\;\;\;
        =\inf_{t\geq0}\alpha(W(t,x))\e^{t}\limsup_{\varepsilon\rightarrow0^+}\frac{\e^{\varepsilon}-1}{\varepsilon}
     \nonumber\\
    &\;\;\;\;
        =\tilde{W}(x).
\end{flalign}
Let
\begin{flalign}\label{eq:def-U}
    U(x)=1/\tilde{W}(x).
\end{flalign}
By combining (\ref{eq:bd-of-tilde-W}) and (\ref{eq:def-U}), we have
\begin{flalign}\label{eq:sector-U-inside}
    \mu_1(|x|_{\cl(\R^n\backslash\setS)})\leq U(x) \leq \mu_2(|x|_{\cl(\R^n\backslash\setS)}),\;\;\forall x\in\Int(\setS)
\end{flalign}
where
\begin{flalign}
    &\mu_1(s)=\exp\Bigg[-\frac{\chi_1^{-1}(\chi_2(1/s)+c)}{2}\Bigg],\;\;
    \mu_2(s)=\exp\Bigg[-\frac{1}{2}\chi_1^{-1}\Bigg(\frac{1}{2s}\Bigg)\Bigg].&
     \nonumber
\end{flalign}
Combining (\ref{eq:LgTildeW}) and (\ref{eq:def-U}),
\begin{equation}
    \tilde{W}(x)\nabla{U}(x)g(x,d)
    =-U(x)\nabla{\tilde{W}}(x)g(x,d)
    \geq-U(x){\tilde{W}}(x).
\end{equation}
Thus,
\begin{flalign}\label{eq:deri-U-inside}
    \nabla{U}(x)g(x,d)
    \geq-\mu_3(U(x)),\;\;\forall x\in\Int(\setS)
\end{flalign}
with $\mu_3(s)=s$.

On the other hand, because the auxiliary system (\ref{eq:iss-aux-system}) is GAS with respect to the safe set $\setS$, it follows from \cite[Theorem 2.8]{lin1996smooth} that there is a continuously differentiable Lyapunov function $V:\R^n\rightarrow\R_{\geq0}$ such that, for all $x\in\R^n\backslash\setS$,
\begin{flalign}
    &\mu_4(|x|_\setS)\leq V(x) \leq \mu_5(|x|_\setS),
     \label{eq:bound-converse-lya}\\
    &L_g V(x)\leq-\mu_6(V(x)),
     \label{eq:cover-lya-deriv}
\end{flalign}
where $\mu_4, \mu_5,\mu_6$ are $K_\infty$-functions.

Let
\begin{flalign}\label{eq:hat-h}
     h(x)=
    \left\{
      \begin{array}{ll}
        U(x), & x\in\Int(\setS); \\
        0, &  x\in\partial\setS; \\
        -V(x), & x\in\R^n\backslash\setS.
      \end{array}
    \right.
\end{flalign}
Due to (\ref{eq:sector-U-inside}) and (\ref{eq:bound-converse-lya}), $h(x)$ satisfies (\ref{eq:safety-and-bf-amespaper}). Moreover, combining (\ref{eq:deri-U-inside}) and (\ref{eq:cover-lya-deriv}), we obtain (\ref{eq:robust-ZBF}).

\emph{Case II: Finite-Time Reachability.}
Let $T(x)$ be the first time that the state trajectory $\phi(t,x,d)$ reaches $\partial\setS$, namely,
\begin{flalign}\label{eq:reach-time}
    T(x)=\inf\{\tau\geq0:\phi(\tau,x,d)\in\partial\setS\},\;\;\forall x\in\Int(\setS),\;\forall d\in M_D.
\end{flalign}
Clearly, $T(x)$ is zero on $\partial\setS$ and continuous on $\setS$. Given arbitrary $x\in\Int(\setS)$ and $d\in M_D$, by the definition of $T(x)$ in (\ref{eq:reach-time}) and the continuity of $\phi(t,x,d)$, we have
\begin{flalign}
    T(\phi(t,x,d))
    &=\inf\{\tau\geq0:\phi(\tau,\phi(t,x,d),d)\in\partial\setS\}
        \nonumber\\
    &=\inf\{\tau\geq0:\phi(\tau+t,x,\tilde{d})\in\partial\setS\}
\end{flalign}
where $\tilde{d}$ is a signal such that
\begin{flalign}
    \tilde{d}(s)
    =\left\{
       \begin{array}{ll}
         d(s), & 0\leq s \leq t \\
         d(s-t), & s >t
       \end{array}
     \right.
\end{flalign}
Because $\tilde{d}$ also belongs to $M_D$,
\begin{flalign}\label{eq:relation-settling-T}
    T(\phi(t,x,d))=T(x)-t,\;\;\forall x\in \Int(\setS),\;\forall d\in M_D.
\end{flalign}
Let
\begin{flalign}\label{eq:zbf-def-in-safe-set}
    U(x)=(T(x))^{\frac{1}{1-c}},\;\;x\in\Int(\setS)
\end{flalign}
where $c\in(0,1)$ is a constant. According to \cite[Lemma 4.3]{khalil2002nonlinear}, because $T(x)$ vanishes on $\partial\setS$ and is positive for all $x\in\Int(\setS)$, there exist $K$-functions $\mu_1$ and $\mu_2$ such that (\ref{eq:sector-U-inside}) holds. Additionally, combining (\ref{eq:relation-settling-T}) and (\ref{eq:zbf-def-in-safe-set}),
\begin{flalign}
    \nabla U(x)g(x,d)
    &=\lim_{\varepsilon\rightarrow0^+}\frac{U(\phi(\varepsilon,x,d))-U(x)}{\varepsilon}
     \nonumber\\
    &=\frac{1}{1-c}(T(x))^{\frac{c}{1-c}}\lim_{\varepsilon\rightarrow0^+}\frac{T(\phi(\varepsilon,x,d))-T(x)}{\varepsilon}
     \nonumber\\
    &=-\frac{1}{1-c}(T(x))^{\frac{c}{1-c}},\;\;\;\;\forall x\in\Int(\setS)
\end{flalign}
which is identical to (\ref{eq:deri-U-inside}). Hence, with a similar argument in Case I, it follows that $h(x)$ satisfies (\ref{eq:robust-ZBF}).

\section*{Appendix III: Properties of Legendre-Fenchel Transform}
\setcounter{subsection}{0}

We introduce some useful properties of the Legendre-Fenchel transform.

\begin{lemma}\label{lemma:property-LF}
Let $\gamma$ be a continuously differentiable $K_\infty$-function whose derivative $\gamma'$ is also of class $K_\infty$. Then the following properties hold:
\begin{basedescript}{\desclabelstyle{\pushlabel}\desclabelwidth{0.7cm}}
    \item[\hspace{0.17cm}(i)] $\ell\gamma(r)=r(\gamma')^{-1}(r)-\gamma((\gamma')^{-1}(r))$;
    \item[\hspace{0.17cm}(ii)] $\ell\ell\gamma=\gamma$;
    \item[\hspace{0.17cm}(iii)] $\ell\gamma$ is a $K_\infty$-function;
    \item[\hspace{0.17cm}(iv)] $\ell(a\gamma)(r)=a\ell\gamma(r/a)$ for any $a>0$.
\end{basedescript}
\end{lemma}

\noindent
\textbf{Proof.}
Properties (i)-(iii) have been shown in \cite[Lemma A1]{krstic1998inverse}, while property (iv) follows from $\ell{a\gamma}(r)
    =\int_0^r(a\gamma')^{-1}(s)ds
     =\int_0^r(\gamma')^{-1}(s/a)ds
     =a\int_0^{r/a}(\gamma')^{-1}(s)ds
     =a\ell\gamma(r/a)$.
\hfill $\Box$
\vskip5pt

\begin{lemma}[\cite{krstic1998inverse}]\label{eq:yong-inequality}
For any two vector $x,y$,
\begin{flalign}
    x^\T y\leq\gamma(|x|)+\ell\gamma(|y|)
\end{flalign}
where the equality holds if and only if
\begin{flalign}
    y=\gamma'(|x|)\frac{x}{|x|},\;\;\text{namely},\;\;x=(\gamma')^{-1}(|y|)\frac{y}{|y|}.
\end{flalign}
\end{lemma}


\ifCLASSOPTIONcaptionsoff
  \newpage
\fi



\end{document}